\newif\ifdraft
\draftfalse
\newif\ifcameraready
\camerareadytrue

\documentclass[camera,letterpaper,nomarginnotes,nonarrowgutter]{jpaper}
\newcounter{version}

\PassOptionsToPackage{hyphens}{url}
\usepackage[bookmarks=true,breaklinks=true,colorlinks,linkcolor=black,citecolor=blue,urlcolor=black]{hyperref}

\usepackage[noadjust]{cite}
\usepackage{amsmath,amssymb,amsfonts}
\usepackage{algorithmic}
\usepackage{graphicx}
\usepackage{textcomp}
\usepackage{xcolor}
\usepackage[hyphens]{url}
\usepackage{fancyhdr}

\usepackage{mathptmx} 
\usepackage{setspace}

\usepackage{balance}

\usepackage{footnote}

\usepackage{xspace}
\usepackage[normalem]{ulem}

\usepackage[skip=2pt]{caption}

\usepackage{enumitem}
\usepackage{xcolor}
\usepackage[us,12hr]{datetime}
\usepackage[en-GB, useregional=numeric]{datetime2}

\usepackage{multirow}

\usepackage{booktabs}
\usepackage{glossaries} 
\usepackage{listings}
\usepackage[linesnumbered]{algorithm2e}
\usepackage{siunitx} 
\usepackage{subcaption}
\usepackage{tikz}
\usepackage{multicol}
\usepackage{threeparttable}
\usepackage{titlesec}
\usepackage{authblk}

\definecolor{gfored}{rgb}{0.580, 0.050, 0.211}
\definecolor{ao}{rgb}{0.007, 0.520, 0.867}
\definecolor{moegi}{rgb}{0.357, 0.537, 0.188}
\definecolor{jl}{rgb}{1.0, 0.2, 0.8}
\definecolor{brown(web)}{rgb}{0.65, 0.16, 0.16}
\definecolor{bisque}{rgb}{1.0, 0.89, 0.77}
\definecolor{nbs}{rgb}{0.88, 0.07, 0.37}
\definecolor{yt}{rgb}{0.58, 0.44, 0.86}
\definecolor{iy}{rgb}{0.0, 0.36, 0.05}
\definecolor{burntorange}{rgb}{0.8, 0.33, 0.0}
\usetikzlibrary{fit}
\newcommand*\circledtest[1]{%
  \begin{tikzpicture}[baseline=(char.base), 
      circ/.style={shape=circle,draw,fill,inner sep=0.3pt}]
      \node[circ] (char) at (3ex,0) {\textcolor{white}{#1}};
  \end{tikzpicture}%
}

\usepackage{tikz}
\usetikzlibrary{calc}
\newcommand*\circledt[2]{\tikz[baseline=(char.base)]{
    \node[shape=circle, draw, fill=#1, inner sep=0.05pt] (char) {\vphantom{WAH1g}\textcolor{white}{#2}};}}

\newcommand{\dingOne}{\circledtest{1}}
\newcommand{\dingTwo}{\circledtest{2}}
\newcommand{\dingThree}{\circledtest{3}}
\newcommand{\dingFour}{\circledtest{4}}
\newcommand{\dingFive}{\circledtest{5}}

\newcommand{\dingA}{\circledt{blue}{a}}
\newcommand{\dingB}{\circledt{blue}{b}}
\newcommand{\dingC}{\circledt{blue}{c}}
\newcommand{\dingD}{\circledt{blue}{d}}
\newcommand{\dingE}{\circledt{blue}{e}}
\newcommand{\dingF}{\circledt{blue}{f}}
\newcommand{\dingG}{\circledt{blue}{g}}

\newcommand{\one}{1)}
\newcommand{\two}{2)}
\newcommand{\three}{3)}
\newcommand{\four}{4)}

\newcommand{\rf}[0]{\texttt{R$_{\texttt{F}}$}}
\newcommand{\nrf}[0]{\texttt{N$_{\texttt{RF}}$}}
\newcommand{\rl}[0]{\texttt{R$_{\texttt{L}}$}}
\newcommand{\nrl}[0]{\texttt{N$_{\texttt{RL}}$}}

\newcommand{\apaSub}[0]{{\texttt{ACT}~\rf{}~$\rightarrow$~\texttt{PRE}~$\rightarrow$~\texttt{ACT}~\rl{}}}
\newcommand{\apaEx}[0]{{\texttt{ACT~src}~$\rightarrow$~\texttt{PRE}~$\rightarrow$~\texttt{ACT~dst}}}
\newcommand{\apaLong}[0]{{\texttt{ACT}~$\rightarrow$~\texttt{PRE}~$\rightarrow$~\texttt{ACT}}}
\newcommand{\src}[0]{{\texttt{src}}}
\newcommand{\dst}[0]{{\texttt{dst}}}
\newcommand{\apa}[0]{\texttt{APA}}

\newcommand{\rref}[0]{\texttt{R$_{\texttt{REF}}$}}
\newcommand{\rcom}[0]{\texttt{R$_{\texttt{COM}}$}}
\newcommand{\apaAnd}[0]{{\texttt{ACT}~\rref{}~$\rightarrow$~\texttt{PRE}~$\rightarrow$~\texttt{ACT}~\rcom{}}}

\newcommand{\rrefx}[1]{\texttt{R$_{\texttt{REF$_{\texttt{#1}}$}}$}}
\newcommand{\rcomx}[1]{\texttt{R$_{\texttt{COM$_{\texttt{#1}}$}}$}}

\newcommand{\vref}[0]{\texttt{V$_{\texttt{REF}}$}}
\newcommand{\vcom}[0]{\texttt{V$_{\texttt{COM}}$}}

\newcommand{\vand}[0]{\texttt{V$_{\texttt{AND}}$}}
\newcommand{\vor}[0]{\texttt{V$_{\texttt{OR}}$}}

\newcommand{\xxx}[1]{\param{XXX}} 

\newcommand{\ignore}[1]{}

\ifdraft
    \usepackage[colorinlistoftodos,prependcaption,textsize=tiny]{todonotes}
    \usepackage{color,soul}
    \paperwidth=\dimexpr \paperwidth + 4cm\relax
    \oddsidemargin=\dimexpr\oddsidemargin + 2cm\relax
    \evensidemargin=\dimexpr\evensidemargin + 2cm\relax
    \marginparwidth=\dimexpr \marginparwidth + 2cm\relax
    \newcommand{\param}[1]{\textcolor{red}{#1}} 
    
    \newcommand{\agycomment}[1]{\todo[size=\scriptsize, linecolor=orange, bordercolor=orange, backgroundcolor=white]{\textcolor{gfored}{\textbf{@gy:} #1}}}
    \renewcommand{\agycomment}[1]{{\textcolor{gfored}{\textbf{\hl{[@gy:}} \hl{#1}\textbf{]}}}}

    \newcommand{\mscomment}[1]{\todo[size=\scriptsize, linecolor=orange, bordercolor=orange, backgroundcolor=white]{\textcolor{red}{\textbf{@MS:} #1}}}
    \renewcommand{\mscomment}[1]{{\textcolor{red}{\textbf{\hl{[@MS:}} \hl{#1}\textbf{]}}}}

    \newcommand{\atb}[1]{\textcolor{ao}{#1}}

    \newcommand{\yctcomment}[1]{\todo[size=\scriptsize, linecolor=orange, bordercolor=orange, backgroundcolor=white]{\textcolor{yt}{\textbf{@yct:} #1}}}
    \renewcommand{\yctcomment}[1]{{\textcolor{yt}{\textbf{\hl{[@yct:}} \hl{#1}\textbf{]}}}}

    \newcommand{\gfcomment}[1]{\todo[size=\scriptsize, linecolor=orange, bordercolor=orange, backgroundcolor=white]{\textcolor{blue}{\textbf{@gf:} #1}}}

    \newcommand{\nbcomment}[1]{\todo[size=\scriptsize, linecolor=orange, bordercolor=orange, backgroundcolor=white]{\textcolor{nbs}{\textbf{@nb:} #1}}}

    \newcommand{\hluocomment}[1]{\todo[size=\scriptsize, linecolor=orange, bordercolor=orange, backgroundcolor=white]{\textcolor{moegi}{\textbf{@hluo:} #1}}}

    \newcommand{\iey}[1]{\textcolor{iy}{#1}}
    \newcommand{\ieyinline}[1]{{\color{iy}{\textbf{\hl{[@iey:}} \textit{\hl{#1}}\textbf{]}}}}
    \newcommand{\ieycomment}[1]{\todo[size=\scriptsize, linecolor=orange, bordercolor=orange, backgroundcolor=white]{\textcolor{iy}{\textbf{@iey:} #1}}}

    \newcommand{\om}[1]{\textcolor{iy}{#1}}
    \newcommand{\omcomment}[1]{\todo[size=\scriptsize, linecolor=orange, bordercolor=orange, backgroundcolor=white]{\textcolor{teal}{\textbf{@om:} #1}}}
    \newcommand{\ominline}[1]{{\textcolor{teal}{\textbf{[@om:} #1\textbf{]}}}}

\else
    \usepackage{soul}
    
    \newcommand{\param}[1]{\textcolor{black}{#1}} 
    
    \newcommand{\agycomment}[1]{}
    \newcommand{\agyinline}[1]{}

    \newcommand{\mscomment}[1]{}
    
    \newcommand{\atb}[2]{\ifnum#1=\value{version}\textcolor{ao}{#2}\else{#2}\fi}

    \newcommand{\yctcomment}[1]{}

    \newcommand{\gfcomment}[1]{}

    \newcommand{\nbcomment}[1]{}

    \newcommand{\hluocomment}[1]{}

    \newcommand{\iey}[2]{\ifnum#1=\value{version}\textcolor{blue}{#2}\else{#2}\fi}

    \newcommand{\ieycomment}[1]{\todo[size=\tiny, linecolor=orange, bordercolor=orange, backgroundcolor=white]{\textcolor{iy}{\textbf{@Ismail:} #1}}}

    \newcommand{\ieyinline}[1]{}

    \newcommand{\om}[2]{\ifnum#1=\value{version}\textcolor{blue}{#2}\else{#2}\fi}
    \newcommand{\omcomment}[1]{\todo[size=\tiny, linecolor=orange, bordercolor=orange, backgroundcolor=white]{\textcolor{red}{\textbf{@Onur:} #1}}}
    \newcommand{\ominline}[1]{}

\fi

\definecolor{frenchblue}{rgb}{0.19, 0.55, 0.91}

\usepackage[most]{tcolorbox} 
\tcbset{before skip=1.5pt, after skip=4pt}

\newtcolorbox[auto counter]{obsx}[3][]{%
    colframe = #2!45,
    colback  = #2!10,
    coltitle = #2!20!black, 
    colbacktitle=#2!20,
    coltitle=black,
    fonttitle=\bfseries, 
    title=#3~\thetcbcounter.\ ,
    enhanced,
    attach boxed title to top left={yshift=-2.8mm, xshift=0.15cm},
    bottom=-2.2pt,
    #1%
}

\usepackage[most]{tcolorbox} 
\newtcolorbox[auto counter]{tkx}[2][]{%
    enhanced, breakable, center title,
    colframe = #2!45,
    colback  = #2!10,
    colbacktitle=#2!20,
    left=-0.5pt,
    right=-0.5pt,
    bottom=-2pt,
    top=-0.25pt,
    #1%
}
\newcounter{obs}
\newcommand\observation[1]{
\refstepcounter{obs}
\begin{tkx}{moegi}
\noindent\textbf{Observation~\theobs.} #1
\end{tkx}
}

\newcounter{tkw}
\newcommand\takeaway[1]{
\stepcounter{tkw}
\begin{tkx}{frenchblue}
\noindent\textbf{Takeaway~\thetkw.} #1
\end{tkx}
}

\newcommand{\PUDAllCitations}[0]{\cite{angizi2019graphide, bostanci2022drstrange,ferreira2022pluto,Li2018SCOPEAS,
olgun2021quactrng, gao2022frac, kim2019drange, kim2018dram,olgun2023dram,olgun2022pidram,
besta2021sisa,deng2018dracc,gao2019computedram,hajinazar2020simdram,li2017drisa,oliveira2022accelerating,seshadri2016buddy,seshadri2013rowclone,seshadri2015fast,seshadri2016processing,seshadri2017ambit,seshadri.bookchapter17,seshadri2018rowclone,seshadri2019dram,xin2020elp2im}}

\newcommand{\dataMovementProblemsCitations}[0]{\cite{
mutlu2013memory,
mutlu2015research,
dean2013tail,
kanev_isca2015,
ferdman2012clearing,
wang2014bigdatabench,
mutlu2019enabling,
mutlu2019processing,
mutlu2020intelligent,
ghose.ibmjrd19,
mutlu2020modern,
boroumand2018google, 
boroumand2021google,
wang2016reducing, 
pandiyan2014quantifying,
koppula2019eden,
kang2014co,
mckee2004reflections,
wilkes2001memory,
kim2012case,
wulf1995hitting,
ghose.sigmetrics20,
ahn2015scalable,
PEI,
hsieh2016transparent,
wang2020figaro,
sites1996,
deoliveira2021IEEE}}

\newcommand\vivekpud{\cite{seshadri2019dram, seshadri2017ambit, seshadri.bookchapter17, seshadri2013rowclone,seshadri2015fast,seshadri2016buddy, seshadri2016processing,seshadri2018rowclone}}

\newcommand\drampumbackground{\cite{hassan2019crow, ghose.sigmetrics20, ghose2018your, kim2016ramulator, seshadri2019dram, kim2012case, zhang2014half, hassan2016chargecache, lee2013tiered, seshadri2017ambit, chang2017understanding,
chang2017understandingphd,chang.sigmetrics2016, chang2014improving, chang2016low, lee2015adaptive, lee2016reducing, lee2016reducingthesis, lee2015decoupled, liu2013experimental, liu2012raidr, seshadri2015gather, ipek2008self, lee2016simultaneous, dennard1968dram, keeth2007dram, mineshphd, hasanphd, o2021energy,
angizi2019graphide,besta2021sisa,bostanci2022drstrange,deng2018dracc,ferreira2022pluto,gao2019computedram,li2017drisa,Li2018SCOPEAS,olgun2021quactrng,olgun2022pidram,seshadri.bookchapter17, seshadri2013rowclone,seshadri2015fast,seshadri2016buddy, seshadri2016processing,seshadri2018rowclone, xin2020elp2im}}

\newcommand{\figref}[1]{Fig.~\ref{#1}}

\newcommand{\secref}[1]{§\ref{#1}}

\makeatletter
\g@addto@macro{\normalsize}{%
  \setlength{\abovedisplayskip}{2pt plus 1pt minus 1pt}
  \setlength{\belowdisplayskip}{2pt plus 1pt minus 1pt}
  \setlength{\abovedisplayshortskip}{0pt}
  \setlength{\belowdisplayshortskip}{0pt}
  \setlength{\intextsep}{2pt plus 1pt minus 1pt}
  \setlength{\textfloatsep}{4pt plus 1pt minus 1pt}
  \setlength{\skip\footins}{5pt plus 1pt minus 1pt}}
\makeatother

\definecolor{amber}{rgb}{1.0, 0.49, 0.0}
\definecolor{bronze}{rgb}{0.8, 0.5, 0.2}
\ifcameraready
    \renewcommand{\atb}[2]{\ifnum#1=99\textcolor{ao}{#2}\else{#2}\fi}
    \renewcommand{\iey}[2]{\ifnum#1=99\textcolor{blue}{#2}\else{#2}\fi}
    \renewcommand{\om}[2]{\ifnum#1=99\textcolor{blue}{#2}\else{#2}\fi}
    \renewcommand{\omcomment}[1]{}
    \renewcommand{\ieycomment}[1]{}

\fi

\newcommand{\nChip}[0]{256}
\newcommand{\nModule}[0]{22}

\newcommand{\tras}[0]{$t_{RAS}$}
\newcommand{\trp}[0]{$t_{RP}$}
\newcommand{\trc}[0]{t_{RC}}
\newcommand{\trefi}[0]{t_{REFI}}
\newcommand{\trefw}[0]{t_{REFW}}
\newcommand{\trfc}[0]{t_{RFC}}
\newcommand{\trrd}[0]{t_{RRD}}

\newcommand{\act}[0]{\texttt{ACT}}
\newcommand{\pre}[0]{\texttt{PRE}}
\newcommand{\refresh}[0]{REF}
\newcommand{\wri}[0]{\texttt{WR}}
\newcommand{\rd}[0]{\texttt{RD}}

\newcommand{\vddh}[0]{\texttt{VDD/2}}

\newcommand{\pum}[0]{\texttt{PuM}}

\usepackage[shortcuts]{extdash}

\newcommand{\versionnum}[0]{5.0}

\ifcameraready
    \fancypagestyle{plain}{
    \fancyhead{}
    }
\else
    \fancypagestyle{firstpage}
    {
    }
\fi

\makeatletter
\def\bstctlcite{\@ifnextchar[{\@bstctlcite}{\@bstctlcite[@auxout]}}
\def\@bstctlcite[#1]#2{\@bsphack
  \@for\@citeb:=#2\do{%
    \edef\@citeb{\expandafter\@firstofone\@citeb}%
    \if@filesw\immediate\write\csname #1\endcsname{\string\citation{\@citeb}}\fi}%
  \@esphack}
\makeatother 

\begin{document}
\bstctlcite{IEEEexample:BSTcontrol}

\title{Functionally-Complete Boolean Logic in Real DRAM Chips: \\ Experimental Characterization and Analysis}
\renewcommand\Authsep{\quad}
\renewcommand*{\Authands}{\quad}

\author{{\.I}smail~Emir~Y{\"u}ksel}
\author{Yahya~Can~Tu\u{g}rul}
\author{Ataberk~Olgun}
\author{F.~Nisa~Bostanc{\i}}
\author{A.~Giray~Ya\u{g}l{\i}k\c{c}{\i}}
\author{Geraldo~F.~Oliveira}
\author{Haocong~Luo} 
\author{Juan~Gómez-Luna}
\author{Mohammad~Sadrosadati}
\author{Onur~Mutlu}
\affil{ETH~Z{\"u}rich}
\date{}   
\ifcameraready
\else
  \thispagestyle{firstpage}
\fi
\pagestyle{plain}

\maketitle

\setcounter{version}{3}
\begin{abstract}
Processing-using-DRAM (PuD) is an emerging paradigm that leverages the analog \om{0}{operational} properties of DRAM circuitry to enable massively parallel in-DRAM computation. PuD has the potential to significantly reduce or eliminate costly data movement between processing elements and main memory. A common approach for PuD architectures is to make use of bulk bitwise computation (e.g., AND, OR, NOT). Prior works experimentally demonstrate \iey{0}{three-input MAJ (i.e., MAJ3)} \om{3}{and} two-input AND \om{0}{and} OR operations in \om{1}{commercial} off-the-shelf (COTS) DRAM chips. \om{0}{Yet, demonstrations on \iey{0}{COTS} DRAM chips do \om{2}{\emph{not}} provide a functionally complete set of operations (e.g., NAND or AND and NOT).}

We experimentally demonstrate that \iey{0}{COTS} DRAM chips are capable of performing 1) \emph{functionally-complete} Boolean operations: NOT, NAND, and NOR and 2) many-input (i.e., more than two\om{2}{-}input) AND and OR operations.
We present an extensive characterization of \emph{new} bulk bitwise operations in \nChip{} off-the-shelf modern DDR4 DRAM chips. We evaluate the reliability of these operations using a metric called \iey{1}{success rate}:~the \om{2}{fraction} of correctly performed bitwise operations. Among our \om{3}{19} new observations, we highlight \iey{1}{four} \om{0}{major} results. First, we can perform the NOT operation on \iey{0}{COTS} DRAM chips with a 98.37\% \iey{1}{success rate} on average. Second, \iey{0}{we can} perform up to 16-input NAND, NOR, AND, and OR operations on \iey{0}{COTS DRAM chips} with high reliability (e.g., 16-input NAND, NOR, AND, and OR with an average \iey{1}{success rate} of~ 94.94\%, 95.87\%, 94.94\%, and 95.85\%, respectively). \iey{1}{Third, data pattern \om{2}{only} slightly affects NAND, NOR, AND, and OR operations.~Our results show that executing NAND, NOR, AND, and OR operations with random data patterns decreases the \iey{1}{success rate} compared to all logic-1/logic-0 patterns by 1.39\%, 1.97\%, 1.43\%, and 1.98\%, respectively. Fourth}, NOT, NAND, NOR, AND, and OR operations are \iey{3}{highly} resilient to temperature changes, with \om{1}{small} \iey{1}{success rate} fluctuations of at most 1.66\% among all the tested operations when the temperature is increased from 50$^{\circ}$C to 95$^{\circ}$C. \om{0}{We believe these \om{3}{empirical} results demonstrate the \om{3}{promising} potential of using DRAM as a computation substrate.} To aid future research and development, we open-source our infrastructure at \url{https://github.com/CMU-SAFARI/FCDRAM}
\end{abstract}

\glsresetall{}
\setcounter{version}{3}
\section{Introduction}
\label{sec:introduction}

Modern systems \om{2}{are processor-centric\om{2}{~\cite{mutlu2019processing,mutlu2020modern}}: they} require frequent data movement between processing elements (e.g., CPU, GPU, \om{2}{TPU, and FPGA}) and main memory~\om{1}{(DRAM)}, leading to significant inefficiencies in performance and energy consumption~\dataMovementProblemsCitations{}.~\om{2}{D}\iey{1}{ata movement from/to main memory is an increasingly significant bottleneck across a wide variety of computing systems and applications}\om{2}{~\cite{boroumand2021google,boroumand2018google}.}~\om{1}{P}rocessing-using-DRAM (PuD)\om{2}{~\cite{seshadri2017ambit,seshadri2015fast,seshadri2019dram,hajinazar2020simdram}} is a promising paradigm \om{1}{that can alleviate the data movement bottleneck}. PuD uses the analog \iey{0}{operational} properties of the DRAM circuitry to enable massively parallel in-DRAM computation. Many prior works~\PUDAllCitations{} demonstrate that PuD can greatly reduce or eliminate data movement.

A widely used approach for PuD is to perform bulk bitwise operations, i.e., bitwise operations on large bit vectors.~\iey{1}{To perform bulk bitwise operations using DRAM}, \om{1}{p}rior works propose modifications to the DRAM circuitry ~\cite{li2017drisa,besta2021sisa,Li2018SCOPEAS,seshadri2016buddy,seshadri2015fast,seshadri2016processing,seshadri2017ambit,seshadri2019dram,wu2022dram,xin2019roc,seshadri.bookchapter17,seshadri2013rowclone,xin2020elp2im,seshadri2018rowclone,ferreira2022pluto,deng2018dracc,angizi2019graphide,deng2019lacc,sutradhar2021look,sutradhar2020ppim}.~Recent works~\cite{gao2019computedram,gao2022frac,olgun2023dram,olgun2022pidram} experimentally demonstrate the feasibility of executing \om{2}{data copy \& initialization~\cite{gao2019computedram,olgun2022pidram}, \om{3}{i.e., the} RowClone operation~\cite{seshadri2013rowclone}, and} \iey{1}{a subset of bitwise operations, i.e., three-input bitwise majority (MAJ3) and two-input AND and OR operations in unmodified commercial off-the-shelf (COTS) DRAM chips by operating beyond manufacturer-recommended DRAM timing parameters. To do so, these works~\cite{gao2019computedram,gao2022frac,olgun2023dram,olgun2022pidram} issue carefully engineered sequence\om{3}{s} of DRAM commands that allow \iey{2}{sequentially} or simultaneously activating multiple DRAM rows. By simultaneously activating multiple DRAM rows, COTS DRAM chips can perform \om{2}{the} MAJ3 operation \om{2}{in a bulk bitwise manner} among activated DRAM rows, which can be used to implement two-input AND and OR operations.}

\iey{1}{While the demonstration of MAJ3 \iey{3}{and} two-input AND and OR operations in COTS DRAM is clearly interesting and shows the potential of investigating PuD as a serious computation substrate, \om{2}{prior works~\cite{gao2019computedram,gao2022frac,olgun2023dram}} do \emph{not} provide the demonstration of 1)~a functionally-complete set of operations (e.g., AND and NOT \om{2}{or NAND/NOR}) \atb{1}{or 2)~AND and OR operations with more than \om{3}{two} inputs (e.g., a 16-input AND operation)}.}

\iey{1}{In this paper,} we experimentally demonstrate that COTS DDR4 DRAM chips are capable of performing \one{}  
NOT, NAND, and NOR operations and \two{} many-input (i.e., \iey{1}{up to 16}\om{2}{-input}) NAND, NOR, AND, and OR operations.
\atb{1}{Doing so allows a COTS DDR4 chip to provide a \om{2}{\emph{functionally-complete}} set of 
many-input }\iey{1}{Boolean operations}.
We present an extensive characterization of the \emph{new} bulk bitwise operations (i.e., NOT, NAND, and NOR) and~\iey{1}{many-input} NAND, NOR, AND, and OR operations \atb{1}{i}n \nChip{} COTS DDR4 chips. We evaluate the reliability of these Boolean operations \atb{1}{using numerous}
data patterns and \atb{1}{at different} DRAM chip temperature\atb{1}{s}. To quantitatively evaluate the reliability \iey{1}{of} \atb{1}{a Boolean operation}, we 
measure \om{2}{its} \om{2}{\emph{success rate}} \iey{2}{per DRAM cell} \atb{1}{as the fraction of} 
\iey{2}{correctly performed bitwise operation\om{3}{s} over \om{3}{10000 trials}}.

\iey{1}{Based on our characterization of NOT and many-input NAND, NOR, AND, and OR operations}, we make \param{19} new \om{2}{empirical} observations and share \om{2}{five} key takeaway lessons from our observations.~\iey{1}{We highlight \param{four} of our major \om{2}{new} \om{2}{results}.}~\iey{1}{First,} off-the-shelf DRAM chips are capable of performing \om{2}{the} NOT operation. We observe that we can perform \om{2}{bulk bitwise} NOT on COTS DRAM chips with an average \iey{1}{success rate} of \param{98.37}\%.\footnote{\iey{2}{We \om{3}{define the} \emph{average} success rate of a bitwise operation as the mean
of all tested DRAM cells’ success rate across all tested DRAM chips.}} \iey{1}{Second,} we can perform NAND, NOR, AND, and OR operations on COTS DRAM chips with $\{2,4,8,16\}$ inputs. Our results show that we can execute 16-input NAND, NOR, AND, and OR operations with an average \iey{1}{success rate} of 94.94\%, 95.87\%, 94.94\%, and 95.85\%, respectively. Third, data pattern affects the \iey{1}{success rate} of \iey{1}{NAND, NOR, AND, and OR} operations\om{2}{, but only slightly.}
We observe that executing NAND, NOR, AND, and  OR operations with random data patterns decreases the \iey{1}{success rate} compared to all logic-1/logic-0 patterns by \param{1.39}\%, \param{1.97}\%, \param{1.43}\%, and \param{1.98}\%, respectively. Fourth, \iey{1}{NOT, NAND, NOR, AND, and OR} operations are \iey{0}{highly} resilient to temperature changes on the DRAM chip. Our results show that an increase in temperature from 50$^{\circ}$C to 95$^{\circ}$C \om{2}{reduces the} \iey{1}{success rate} by \iey{3}{at most only 1.66\%} across all tested bulk bitwise operations.

\om{2}{We} explain how a COTS DRAM chip can perform 1)~NOT and 2)~many-input NAND, NOR, AND, and OR operations \om{2}{by providing} two hypotheses \om{2}{into the underlying operation of COTS DRAM chips}. \atb{1}{We describe these two \iey{2}{hypotheses} \om{3}{in detail in}~\secref{subsec:hypo_not} and~\secref{subsec:hypo_nand}.}

\begin{figure}[htbp]
\centering
\begin{subfigure}[b]{0.475\linewidth}
\centering
\includegraphics[width=\linewidth]{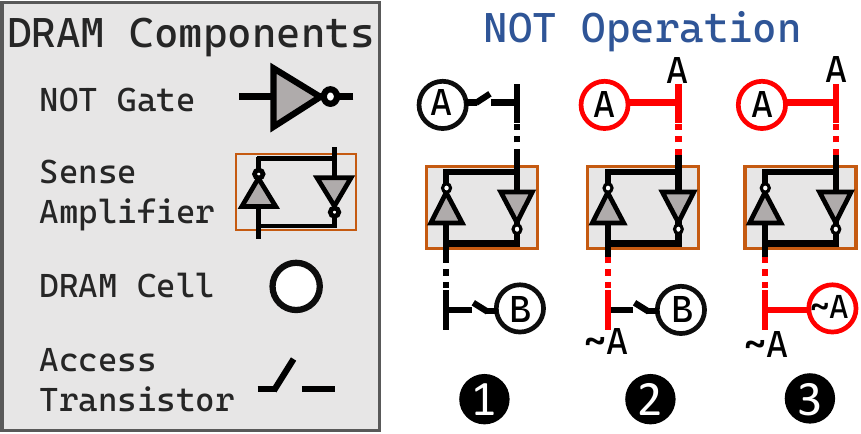}
\caption{}
\label{fig:intro_not}
\end{subfigure}\hspace{12pt}
\begin{subfigure}[b]{0.475\linewidth}
\centering
\includegraphics[width=\linewidth]{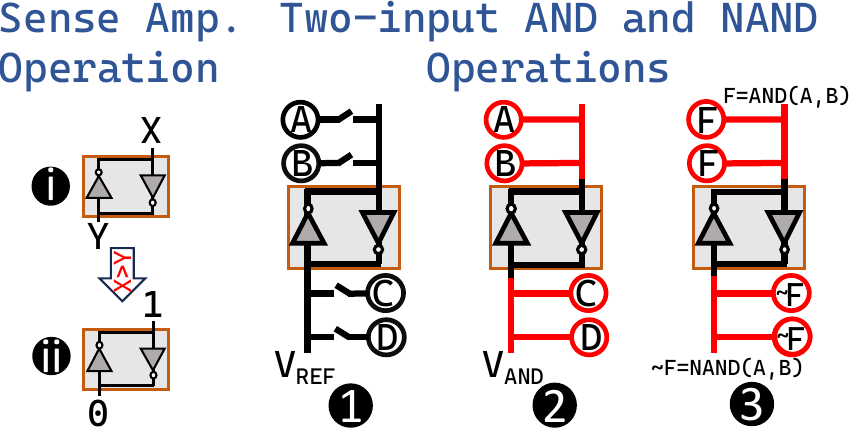}
\caption{}
\label{fig:intro_nand}
\end{subfigure}
\caption{\om{3}{Demonstration of how modern DRAM can provide the NOT operation (a) and two-input AND and NAND operations (b).}}
\label{fig:intro_fig}
\end{figure}

\atb{1}{First, in the widely adopted~\cite{luo2020clrdram,chang2016lisa,sekiguchi2002low,jacob2009memory} open-bitline DRAM architecture~\cite{jacob_book_2008,itoh2013vlsi,dram-circuit-design,jacob2009memory,schloesser20086f}, \emph{two relatively far apart DRAM cells} \iey{2}{(e.g., Cell A and Cell B in \figref{fig:intro_not}-\dingOne{})} connect to \emph{two opposite terminals} of a sense amplifier (\iey{2}{\figref{fig:intro_not}-\dingOne{}}) \iey{3}{via access transistors}. Shortly after a sense amplifier is enabled to access a DRAM cell, the opposite terminals of the sense amplifier are fundamentally driven by inverted voltage levels (i.e., inverted logic values) due to how the sense amplifier operates~\cite{dram-circuit-design,chang2016lisa,jacob_book_2008,jacob2009memory,itoh2013vlsi,weste2015cmos}~\iey{2}{(e.g., A and $\sim$A in \figref{fig:intro_not}-\dingTwo{})}. During standard DRAM operation, \emph{only} one DRAM cell \om{3}{is} connect\om{3}{ed} to one of the terminals, and thus, the other terminal is driven by the inverse of the connected cell's value~\cite{dram-circuit-design,chang2016lisa,jacob_book_2008,jacob2009memory,sekiguchi2002low,itoh2013vlsi,weste2015cmos}. \om{2}{W}e hypothesize that when we \om{3}{\emph{simultaneously connect}} the two DRAM cells to the two sense amplifier terminals (by simultaneously activating the rows of these cells), we can negate the value stored in one cell and store the negated value in the other cell (\iey{2}{\figref{fig:intro_not}-\dingThree{}}).} 

\iey{1}{Second, a sense amplifier \atb{1}{operates in two steps to access a DRAM cell, where it} i)~compares the voltage level\atb{1}{s} of its two opposite terminals \iey{2}{(i.e., \circledt{black}{i} in \figref{fig:intro_nand})} and ii)~amplifies the voltage difference between \atb{1}{them} \iey{2}{(\circledt{black}{ii} in \figref{fig:intro_nand})~\cite{dram-circuit-design,jacob_book_2008,chang2016lisa,kim2019drange,luo2020clrdram,lee2013tiered,khan2016parbor,jacob2009memory,itoh2013vlsi,weste2015cmos}.} \atb{1}{This voltage difference (at the beginning of a DRAM cell access operation) is mainly a function of the value stored in the accessed DRAM cell.} One terminal of the sense amplifier connects to the accessed DRAM cell, while the other terminal \iey{3}{(i.e., the reference terminal)} is driven to a fixed voltage level called the \emph{reference voltage} \iey{2}{(i.e., \texttt{V$_{\texttt{REF}}$} in \figref{fig:intro_nand}-\dingOne{})~\cite{dram-circuit-design,kim2012case,jacob_book_2008,chang2016lisa,jacob2009memory,itoh2013vlsi,weste2015cmos}.} We hypothesize that simultaneously activating multiple DRAM cells connected to both terminals allows us to manipulate the reference voltage value before the sense amplifier amplifies the voltage difference between terminals \iey{2}{(e.g., \texttt{V$_{\texttt{AND}}$} in \figref{fig:intro_nand}-\dingTwo{})}. This manipulation enables the voltage difference between the two sense amplifier terminals to express a wider variety of functions than just the value stored in an accessed DRAM cell.} \iey{3}{By carefully initializing DRAM cells connected to the reference terminal (e.g., Cell C and Cell D in \figref{fig:intro_nand}), we can \atb{3}{set} \vref{} to a \atb{3}{desired} voltage level (e.g., \texttt{V$_{\texttt{AND}}$} in \figref{fig:intro_nand}-\dingTwo{}) that enables us to express two functions: AND and OR operations on the cells connected to the other terminal (e.g., Cell A and Cell B in \figref{fig:intro_nand}-\dingOne{},\dingTwo{}). \atb{3}{The key idea for implementing a Boolean AND operation (which is similar to that of a Boolean OR operation) is that if 1) \texttt{V$_{\texttt{AND}}$} is \emph{lower} than the voltage level created on the other sense amplifier terminal \emph{only when both A and B store a logic-1 value}, the sense amplifier will output a logic-1 (that is, the result of \texttt{AND(logic-1,logic-1)}) and 2) \texttt{V$_{\texttt{AND}}$} is \emph{higher} than the voltage level created on the other sense amplifier terminal \emph{for all other combinations of values stored in A and B}, the sense amplifier will output a logic-0.} Expressing the AND (OR) operation in one terminal results in the NAND (NOR) operation in the other terminal, as two opposite terminals share a connection via the NOT gate. For example, in \figref{fig:intro_nand}-\dingThree{}, one terminal becomes \texttt{F=AND(A,B)} while the other terminal becomes $\sim$\texttt{F=NAND(A,B)}.}

This paper makes the following key contributions:
\begin{itemize}
    \item To our knowledge, this is the first work to experimentally demonstrate that \om{2}{unmodified} off-the-shelf DRAM chips are capable of performing \one{} functionally-complete Boolean operations \om{2}{(i.e., NOT, NAND, and NOR)} and \two{} many-input (i.e., more than two\om{2}{-input}) NAND, NOR, AND and OR operations. 
    \item We extensively characterize \iey{2}{the reliability of} \iey{1}{these previously undemonstrated} bulk bitwise operations on \nChip{} modern DRAM DDR4 chips from \nModule{} DRAM modules. Our results show that \iey{0}{we can} \om{2}{reliably} perform NOT operations and ~$\{2,4,8,16\}$\iey{2}{-input} NAND, NOR, AND, and OR operations \iey{0}{on COTS DRAM chips} \om{2}{at high success rates ($>$94\%)}.
   \item We quantitatively evaluate the impact of two significant factors: \iey{1}{1)~data pattern dependence and 2)~DRAM chip temperature} on the \iey{1}{success rate} of bulk bitwise operations in \iey{0}{DRAM chips}. \iey{1}{Our results show that} \iey{2}{the effect of data pattern dependence and temperature on the average success rate of bulk bitwise operations is small, i.e., \iey{3}{$\leq$}1.98\% and \iey{3}{$\leq$}1.66\%, respectively.}
\end{itemize}

\iey{1}{We believe that our \om{2}{experimental results} fills a \om{2}{large} gap in \atb{1}{the understanding of COTS DRAM chips' computational capabilities.}~\om{2}{By} demonstrat\om{2}{ing} the capability of performing functionally-complete Boolean logic in COTS DRAM chips and characteriz\om{2}{ing} this behavior\om{2}{,} \om{2}{we} demonstrate the potential of using DRAM as a \om{2}{powerful} computation substrate.} \om{3}{To aid future research and development, we open-source our infrastructure at \url{https://github.com/CMU-SAFARI/FCDRAM}.}

\glsresetall{}
\setcounter{version}{2}
\section{Background}
\label{sec:background}
We provide a brief background on DRAM organization, DRAM commands, DRAM operation, and \om{1}{Processing-using-DRAM (PuD)} in commercial off-the-shelf (COTS) chips. For more detailed
background on these, we refer the reader to many prior works~\drampumbackground{}.

\subsection{DRAM Organization \& Operation}
\label{subsec:dram_org}

A modern DRAM system consists of a hierarchy of components: channels, modules, ranks, chips, banks, and subarrays. \iey{0}{In a typical system
configuration, a CPU chip includes a set of memory controllers, where each memory controller interfaces with a DRAM channel to serve memory requests (i.e., reads or writes). A DRAM channel can host multiple DRAM modules, each of which implements a single or multiple DRAM ranks. A DRAM rank consists of multiple DRAM chips that operate in lock-step, i.e., all chips simultaneously perform the same operation.}

\begin{figure}[ht]
\centering
\includegraphics[width=\linewidth]{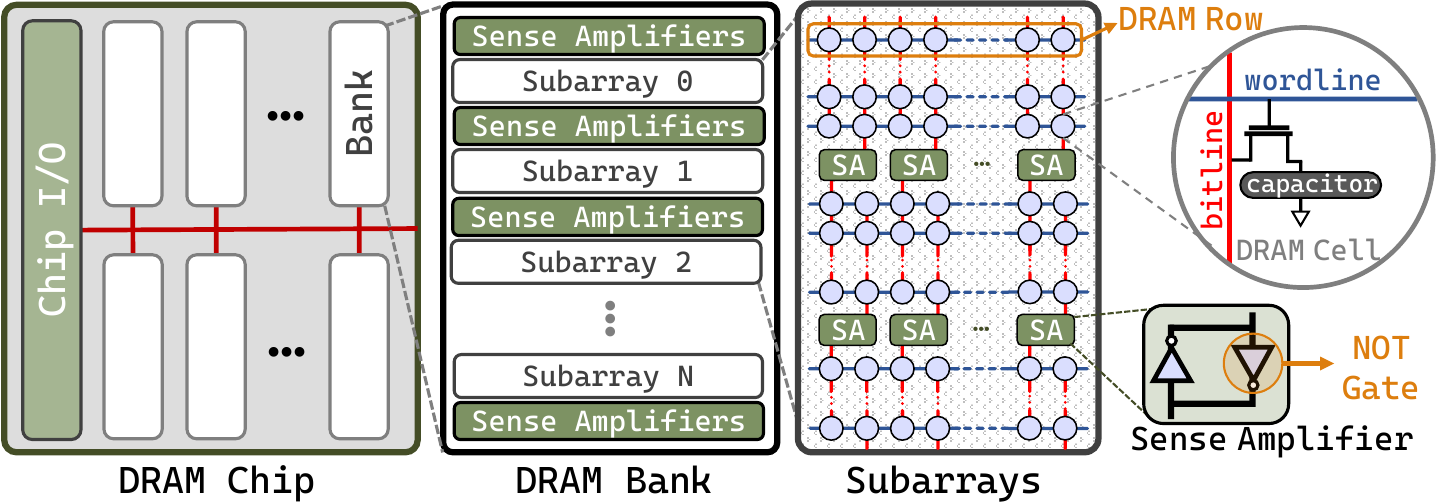}
\caption{\iey{1}{Bank and subarray organization in a DRAM chip.}}
\label{fig:dram_organization}
\end{figure}

\noindent\textbf{DRAM Subarrays.}
\iey{0}{\figref{fig:dram_organization} shows the hierarchical organization of a single DRAM chip that consists of multiple DRAM banks, DRAM subarrays, and DRAM cells.} A \emph{bank} comprises multiple \emph{subarrays}, each containing a two-dimensional array of DRAM cells. A \emph{cell} consists of an access transistor and a storage capacitor that encodes a single bit of data using its voltage level, either the level of supply voltage (VDD) or the level of ground voltage (GND). Cells in a row share a \emph{wordline} that drives each cell's access transistor. DRAM cells in the same column share a \emph{bitline}, which is used to read from and write to the cells via the row buffer (which contains sense amplifiers). A \emph{sense amplifier} forms two back-to-back inverters (i.e., NOT gates) and functions as \om{1}{a} comparator, comparing the voltage of each endpoint of the NOT gate (i.e., bitlines) and amplifying the difference, which results in VDD in the higher voltage point and GND in the lower voltage point. 
However, due to the significant size difference between a sense amplifier and a cell, only enough sense amplifiers are fitted in a row to sense half of the cells~\cite{chang2016lisa,lee2013tiered}. To sense the entire row of cells, each subarray has bitlines that connect to two rows of sense amplifiers, one above and one below the subarray, \om{2}{which causes} neighboring subarrays \om{1}{to} share half of the sense amplifiers. This approach, known as the \om{1}{\emph{open-bitline architecture}}, is widely adopted \om{1}{in} high-density DRAM~\cite{dram-circuit-design,schloesser20086f,sekiguchi2002low,luo2020clrdram,chang2016lisa,jacob_book_2008,itoh2013vlsi,itoh77, lee2013tiered}. For simplicity, we assume that all DRAM cells store VDD for logic-1 and GND for logic-0 in the remainder of this study.

\noindent\textbf{DRAM Commands.}
\iey{0}{To serve main memory requests, the memory controller issues DRAM commands, e.g., row activation (\act{}), bank precharge (\pre{}), data read (\rd{}), data write (\wri{}), and refresh (\texttt{REF}). To perform a read or write operation, the memory controller first needs to open a row, i.e., copy the data of the cells in the row to the row buffer. To open a row, the memory controller issues an \act{} command to a bank by specifying the address of the row to open. After activation completes, the memory controller issues either a \rd{} or a \wri{} command to read or write a DRAM word (which is typically equal to 64 bytes) within the activated row. To access data from another DRAM row in the same bank, the memory controller must first close the currently open row by issuing a \pre{} command. The memory controller also periodically issues \texttt{REF} commands to prevent data loss due to charge leakage.}

\begin{figure}[ht]
\centering
\includegraphics[width=\linewidth]{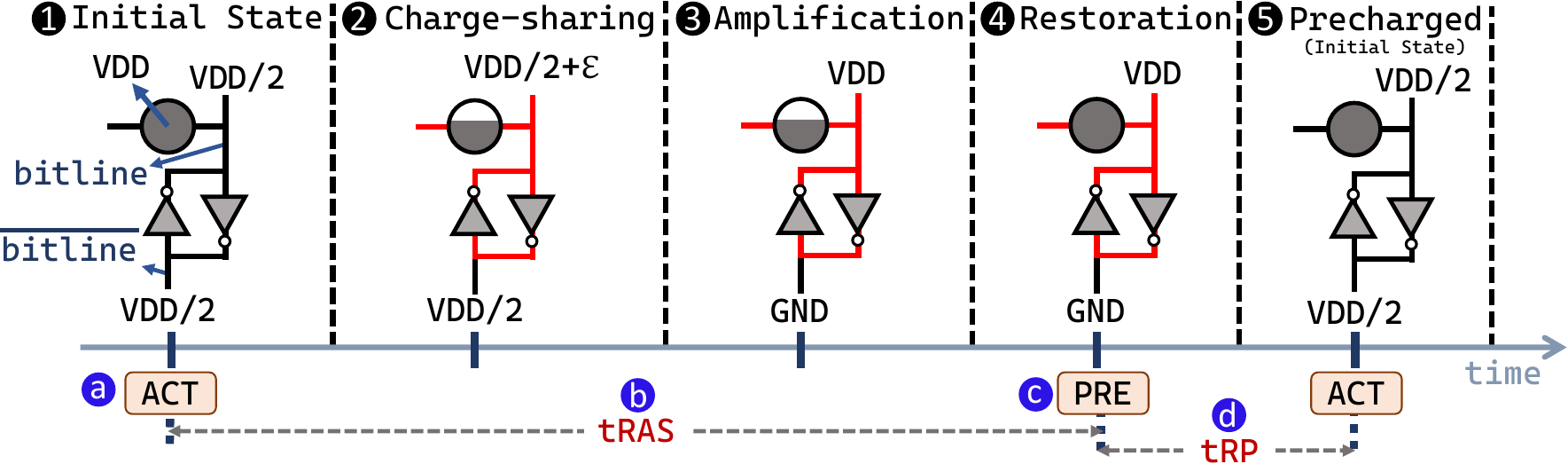}
\caption{\iey{1}{Command sequence for activating a DRAM row and the state of a DRAM cell during each related step.}}
\label{fig:cell_operations}
\end{figure}

\noindent\textbf{DRAM Cell Operation.} We describe DRAM cell operations by explaining the steps in activating a cell. The memory controller initiates each step by issuing a DRAM command. Each step takes a certain amount of time to complete, and thus, a DRAM command is associated with timing constraints known as timing parameters.
In~\figref{fig:cell_operations}, we show how the state of a cell and the sense amplifiers change during the steps involved in an activation operation. Each DRAM cell diagram corresponds to the state of the cell at exactly the tick mark on the time axis, \iey{1}{and asserted signals are highlighted in red.} The memory controller issues each command (shown in orange boxes below the time axis) at the corresponding tick mark. Initially, \one{} cell capacitor stores VDD, \two{} the cell is precharged, \three{} the sense amplifier is disabled, and its bitlines' (i.e., bitline and bitline-bar) voltages are stable at VDD/2 (\dingOne{}).

\om{1}{T}o activate the cell, the memory controller issues an \act{} command (\dingA{}). \om{1}{As a result}, the row decoder asserts the wordline and thus connects the cell capacitor to the bitline. The cell capacitor charge begins to perturb the bitline, known as the \emph{charge-sharing} process (\dingTwo{}). Charge sharing continues until the bitline voltages reach a level that the sense amplifier can safely amplify, VDD/2+$\varepsilon$ (\dingTwo{}). The sense amplifier \om{1}{then kicks in to amplify} the difference between the bitline and the bitline-bar. Depending on the cell's charge, the bitline becomes either VDD or GND, while the bitline-bar becomes the negated voltage value of the bitline. As the cell \om{2}{in \figref{fig:cell_operations}} initially stores VDD, the bitline becomes VDD, whereas the bitline-bar becomes GND at the end of the amplification (\dingThree{}). \om{1}{At this point, the memory controller can read from or write to the cell using \rd{}/\wri{} commands}. To return a cell to its precharged state, the voltage in the cell must first be fully restored, which requires waiting for \tras{} timing parameter (i.e., the timing parameter between \act{} and \pre{} command\om{2}{s}) (\dingB{}). Once the cell is restored, the memory controller can issue a \pre{} command (\dingC{}). The cell returns to the precharge\om{2}{d} state (\dingFive{}), \iey{1}{same as the initial state}\om{2}{,} (\dingOne{}) \om{1}{where the memory controller can reliably issue an \act{} command} after waiting for timing parameter \trp{} (\dingD{}).

\subsection{\iey{0}{Processing Using COTS DRAM Chips}}
\label{subsec:back_pud}
\iey{0}{Processing-using-DRAM (PuD) is an emerging paradigm that can alleviate the bottleneck caused by frequent data movement between processing elements (e.g., CPU) and main memory (DRAM). PuD enables massively parallel in-DRAM computation by leveraging intrinsic analog operational properties of the DRAM circuitry}\om{1}{, as initially shown by~\vivekpud{}.}

\iey{1}{Recent works~\cite{olgun2022pidram,gao2019computedram,gao2022frac,olgun2023dram,olgun2021quactrng} experimentally demonstrate that COTS DRAM chips can perform 1) data copy \& initialization and 2) bitwise operations\om{2}{,} by carefully crafting a set of DRAM commands (i.e., \apaLong{}) with reduced timings (i.e., violating the \tras{} and \trp{} timing parameters).}

\noindent\textbf{\iey{0}{In-DRAM Data Copy \& Initialization.}}
\om{2}{RowClone}\iey{0}{~\cite{seshadri2013rowclone} enables data movement within \om{2}{a} DRAM \om{2}{subarray} \om{2}{at} a DRAM row granularity by modifying DRAM circuitry. RowClone alleviates the energy and execution time costs of transferring data between the DRAM and the processing units. Recent works~\cite{gao2019computedram,olgun2022pidram} experimentally demonstrate that RowClone operation can be performed in COTS DRAM chips by enabling sequential activation of two DRAM rows \om{2}{in the same subarray}.}

\noindent\textbf{\iey{0}{In-DRAM Bitwise Operations.}}
\iey{0}{Prior work~\cite{seshadri2017ambit} introduces \iey{1}{the concept of simultaneously activating three rows in a DRAM subarray (i.e., triple-row activation) through modifications to the DRAM circuitry. This modification enables} \om{2}{a} three-input bitwise majority (MAJ3) operation among activated rows, which can be used to implement two-input bitwise AND and OR operations. \iey{1}{Prior works~\cite{gao2019computedram,olgun2023dram,gao2022frac} demonstrate that COTS DRAM chips are capable of performing MAJ3 and two-input AND and OR operations by simultaneously activating multiple rows in the same subarray.}}

\subsection{\iey{1}{Motivation and Goal}}

\iey{1}{PuD is a promising paradigm that can alleviate the data movement bottleneck and thus improve the overall energy efficiency and performance of DRAM-based systems. To show the potential of using DRAM as a serious computation substrate, it is important to understand and characterize the computational capability of \om{2}{existing} COTS DRAM chips. }

\iey{2}{While the demonstration of MAJ3 \iey{2}{and} two-input AND and OR operations in COTS DRAM is clearly interesting and shows the potential of investigating PuD as a serious computation substrate, \om{2}{prior works~\cite{gao2019computedram,gao2022frac,olgun2023dram}} do \emph{not} provide the demonstration of 1)~a functionally-complete set of operations (e.g., OR and NOT \om{2}{or NAND/NOR}) \atb{1}{or 2)~AND and OR operations with more than \om{2}{two} inputs (e.g., a 16-input OR operation)}.}

Our goal is to 1) understand whether COTS DRAM chips are capable of executing functionally-complete operations and bitwise operations with more than \om{3}{two} inputs and 2) rigorously characterize the reliability of executing bitwise operations in COTS DRAM chips.

\setcounter{version}{2}
\section{Methodology}
\iey{1}{We describe our commercial off-the-shelf (COTS) DRAM testing infrastructure (\secref{subsec:infra}) and the COTS DDR4 DRAM chips tested for our characterization study (\secref{subsec:tested_chips}). We explain the methodology of our \om{2}{different} characterization experiments in their \om{2}{corresponding} sections (\secref{sec:multi_sub},~\secref{sec:not}, and ~\secref{sec:nand}).}

\subsection{\iey{0}{COTS} DRAM Testing Infrastructure}
\label{subsec:infra}
\begin{figure}[ht]
\centering
\includegraphics[width=\linewidth]{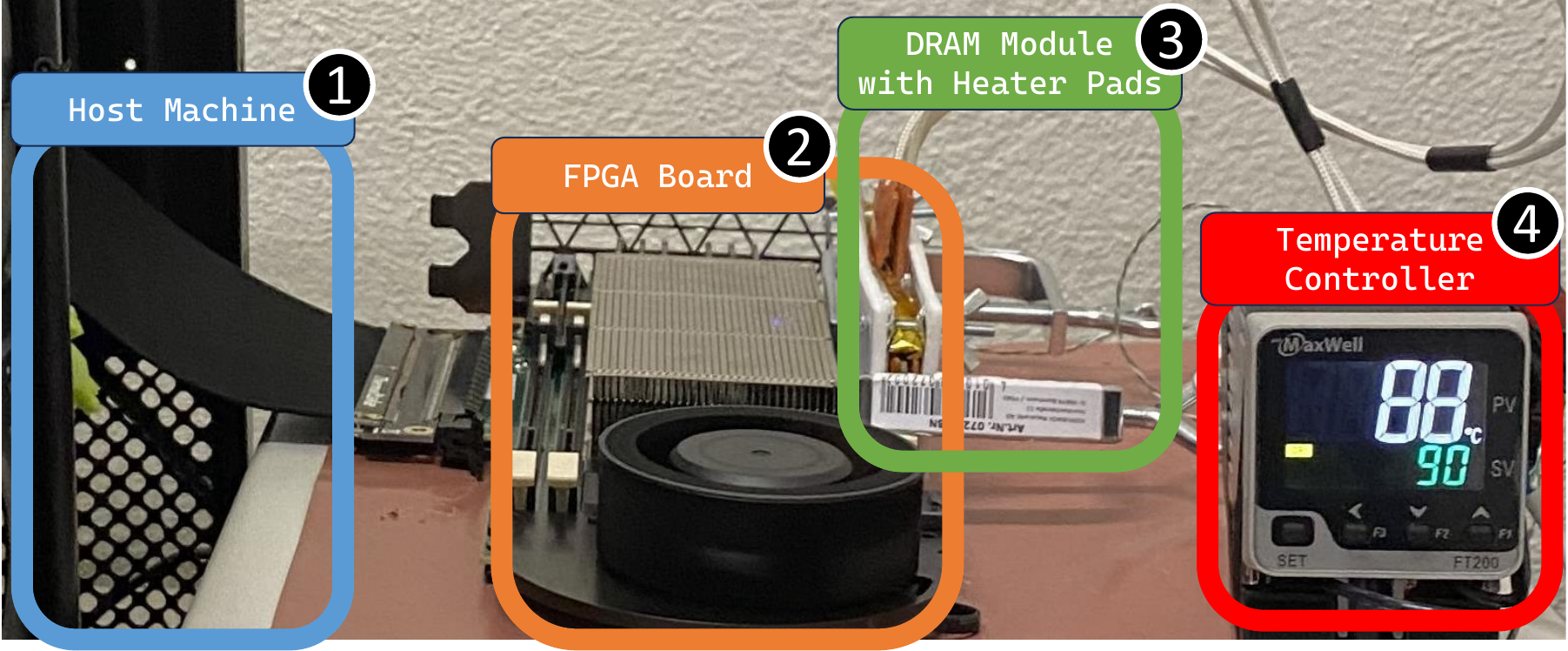}
\caption{\om{2}{Our} DRAM Bender \om{2}{based} experimental setup.}
\label{fig:infra}
\end{figure}

We conduct COTS DRAM chip experiments using DRAM Bender~\cite{safari-drambender, olgun2023dram}, an FPGA-based DDR4 testing infrastructure that provides precise control of DDR4 commands issued to a DRAM module. \figref{fig:infra} shows our experimental setup that consists of four main components: 1) a host machine that generates the test program and collects experiment results, 2) an FPGA development board~\cite{alveo_u200}, programmed with DRAM Bender, 3) a thermocouple temperature sensor and heater pads pressed against the DRAM chips to maintain a target temperature level, and 4) a temperature controller (MaxWell FT200~\cite{maxwellFT200}) that keeps the temperature at the desired level. 

\subsection{\iey{0}{COTS} DDR4 DRAM Chips Tested}
\label{subsec:tested_chips}

Table~\ref{tab:dram_chips} provides \iey{1}{the \nChip{} (\nModule{}) COTS DDR4 DRAM chips (modules)} that we focus our analysis on along with the chip manufacturer (Chip Mfr.), the number of modules (\#Modules), the number of chips (\#Chips), die revision (Die Rev.), DRAM module's manufacturing date (Mfr. Date) in the form of year-week, chip density (Density), chip organization (Org.), and DRAM speed rate in mega transfers per second (MT/s).\footnote{\label{fn:die}{The technology node of a DRAM chip is usually not publicly available. Prior works~\cite{luo2023rowpress,kim2020revisiting,orosa2021deeper}} assume that for a given chip manufacturer and chip density, the alphabetical order of die revision codes may provide an indication of technology node advancement.}

\begin{table}[h!]
\resizebox{\linewidth}{!}{%
\centering
\begin{threeparttable}
\caption{Summary of DDR4 DRAM chips tested.}
\label{tab:dram_chips}
\begin{tabular}{@{}ccccccc@{}}
\toprule
\multirow{2}{*} {{\bf Chip Mfr.}} & \textbf{\#Modules} & {{\bf Die}} & {{\bf Mfr.}} &{{\bf Chip}} & {{\bf Chip}} & {{\bf Speed}} \\ 
& \textbf{(\#Chips)} & {{\bf Rev.}} & {{\bf Date\tnote{a}}} & {{\bf Density}} & {{\bf Org.}} & {{\bf Rate}} \\ 
\midrule
\multirow{6}{*}{SK Hynix}  & 9 (72) & M & N/A & 4Gb & x8 & 2666MT/s\\ 
& 5 (40) & A & N/A & 4Gb & x8 & 2133MT/s \\
 & 1 (16) & A & N/A & 8Gb & x8 & 2666MT/s \\ 
& 1 (32) & A & 18-14 & 4Gb & x4 & 2400MT/s\\
 & 1 (32) & A & 16-49& 8Gb & x4 & 2400MT/s \\
 & 1 (32) & M & 16-22 & 8Gb & x4 & 2666MT/s \\ 
\midrule
\multirow{3}{*}{Samsung} & 1 (8) & F & 21-02 & 4Gb & x8 & 2666MT/s\\ 
 & 2 (16) & D & 21-10 & 8Gb & x8 & 2133MT/s \\
 & 1 (8) & A & 22-12 & 8Gb & x8 & 3200MT/s \\ 
 \bottomrule
\end{tabular}%
\begin{tablenotes}
\item[a] We report “N/A” if no date is marked on the label of a module.
\end{tablenotes}

\end{threeparttable}

}

\end{table}

\iey{1}{To investigate whether our characterization study applies to different DRAM technologies, designs, and manufacturing processes, we test a total of 280 (28) COTS DDR4 DRAM chips (modules) from all three major manufacturers (i.e., SK Hynix, Samsung, and Micron) with different die densities and die revisions from each DRAM chip manufacturer. While we observe successful \iey{0}{NOT, NAND, NOR, AND, and OR operations} in all tested SK Hynix modules, we observe \om{1}{only} successful NOT operations in Samsung chips and no successful bitwise operations in Micron chips. Thus, we focus our analysis on \nChip{} (\nModule{}) COTS DDR4 DRAM chips (modules) from SK Hynix and Samsung, listed in Table~\ref{tab:dram_chips}. We provide much more detail on other chips in the extended version of this paper~\cite{yuksel2024functionally}. We hypothesize why we do not observe all tested bitwise operations in Samsung and Micron chips in~\secref{sec:discussion}.}

\setcounter{version}{3}
\section{Simultaneous Multiple-Row Activation in Neighboring Subarrays in COTS DRAM Chips}
\label{sec:multi_sub}
\atb{1}{We hypothesize that a COTS DRAM chip can perform bitwise NOT and many-input \om{2}{B}oolean operations if such a chip can simultaneously activate multiple DRAM rows in neighboring subarrays, as briefly described in \secref{sec:introduction}. 
For example, activating one row (i.e., \emph{source row}) in one subarray and another row (i.e., \emph{destination row}) in a neighboring subarray in very short succession could negate the value of the source row and store it in the destination row. Thus, comprehensively evaluating the computational capabilities of a COTS DRAM chip necessitates understanding its simultaneous multiple-row activation properties. These properties include, for example, whether or not the DRAM chip can simultaneously activate multiple rows in neighboring subarrays and how many such rows, in each subarray, the chip can simultaneously activate. This section describes our \iey{1}{key idea to perform simultaneous multiple-row activation in neighboring subarrays (\secref{subsec:key_multi_sub})}, our experimental methodology for understanding a DRAM chip's multiple-row activation properties (\secref{subsec:method_multi_sub}) and presents our findings (\secref{subsec:char_multi_sub}).}

\subsection{Multiple-Row Activation in \\Neighboring Subarrays: Key Idea}
\label{subsec:key_multi_sub}
\noindent\textbf{Multiple-Row Activation in One Subarray.}
\iey{1}{
\atb{1}{A carefully} craft\atb{1}{ed} set of \apaSub{} DRAM command sequence with reduced timings (i.e., \atb{1}{with} violat\atb{1}{ed} \tras{} and \trp{} timing parameters) can simultaneously activate multiple rows in the \emph{same subarray}\om{2}{,} \iey{1}{as shown by prior works~\cite{gao2019computedram,gao2022frac,olgun2021quactrng,olgun2023dram,olgun2022pidram}}. In \atb{1}{the} \apaSub{} command sequence, \rf{} and \rl{} denote the row addresses to which the \act{} commands are sent \iey{2}{(i.e., the first \act{} command is sent to \rf{} and the last\om{3}{,} second\om{3}{,} \act{} command is sent to \rl{})}, and \atb{1}{the rows} \rf{} and \rl{} are in the same subarray. 
}

\noindent\textbf{Key Idea.} 
\iey{1}{Our key idea is to simultaneously activate multiple rows in \emph{two neighboring subarrays} by issuing the \apaSub{} command sequence with reduced timings, where \rf{} \atb{1}{points to a row in one} subarray and \rl{} \atb{1}{points to \om{2}{a} row in a neighboring subarray.}
} 

\noindent\textbf{How to Simultaneously Activate Multiple Rows.}
\iey{1}{
Our key idea relies on a fundamental design \om{3}{characteristic} in modern DRAM chips: \om{2}{hierarchical design of the row decoder circuitry}. Modern DRAM chips have multiple \om{2}{levels} of row address decoding stages to reduce latency, area, and power consumption~\cite{bai2022low,weste2015cmos,turi2008high, olgun2021quactrng,dram-circuit-design}. T\om{2}{he hierarchical row decoder structure} expands a row address into a set of intermediate control signals and issuing an \act{} command asserts multiple control signals~\cite{bai2022low,weste2015cmos, olgun2021quactrng,dram-circuit-design}. Prior works~\cite{gao2022frac,olgun2021quactrng} hypothesize that back-to-back \act{} commands with reduced timings assert multiple control signals in the hierarchical row decoder design that \om{2}{can} potentially activate multiple rows. 
}

\noindent\textbf{Key Mechanism.}~\iey{1}{By leveraging the hierarchical structure in the row decoder circuitry, we hypothesize that we can simultaneously activate multiple rows in two neighboring subarrays in three steps. 
First, we issue the \texttt{ACT}~\rf{} command, which activates \rf{} and asserts multiple signals/latches in the hierarchical row decoder circuitry. 
Second, we issue a \texttt{PRE} command with reduced \trp{} timing (e.g., \trp{}$<$3ns), which initiates the de-assertion of signals and latches in the hierarchical row decoder design. 
Third, we issue the \texttt{ACT}~\rl{} command by violating the \tras{} timing (e.g., \tras{}$<$3ns)\om{2}{,} which \one{} prevents the \pre{} command from deasserting the signals \om{2}{or} resetting the latches in the row decoder, and \two{} sets other signals/latches \iey{2}{in the row decoder}. 
}

\subsection{Experimental Methodology}
\label{subsec:method_multi_sub}

\noindent\textbf{DRAM Subarray Boundaries.} \iey{1}{Understanding COTS DRAM chips\atb{1}{'} capability of simultaneously activating multiple rows in neighboring subarrays requires reverse engineering DRAM subarray boundaries.
Prior work~\cite{gao2019computedram,olgun2022pidram} demonstrates that COTS DRAM chips are capable of performing \om{2}{the} RowClone operation~\cite{seshadri2013rowclone}, i.e., copying the content of a DRAM row (i.e., source row) to another DRAM row (i.e., destination row), if the source row and the destination row are in the \emph{same subarray} (see more detail in \secref{subsec:back_pud}). We repeatedly perform the RowClone operation for every possible source and destination row address in each tested bank. When we observe that the destination row \om{2}{gets} the same content as the source row \om{2}{after RowClone}, we conclude that the source row and the destination row are in the same subarray. Based on this observation, we reverse engineer the DRAM subarray boundaries and which rows are in the same subarray.}

\noindent\textbf{Testing Methodology.} \iey{1}{Our experiment consists of three steps. 
First, we initialize the subarrays corresponding to \rf{} and \rl{} with a predefined data pattern.
Second, we issue the \apaSub{} command sequence with reduced timings to simultaneously activate multiple rows in neighboring subarrays (i.e., our key mechanism, described in \secref{subsec:key_multi_sub}).
Third, we issue a \wri{} command with a different data pattern \om{2}{from} the predefined data pattern by respecting the timing parameters. \atb{2}{This \wri{} command cause\om{3}{s} the sense amplifiers to overdrive their bitlines~\cite{dram-circuit-design,marazzi2023rega} and \om{3}{thus} update\om{3}{s} the values of the cells in all simultaneously-activated DRAM rows using the predefined data pattern.}~After the three-step procedure, \atb{2}{to understand which rows from \rf{} and \rl{}'s subarrays are activated by the \apaSub{} command sequence,}
we precharge the bank and read each row in the subarrays of \rf{} and \rl{} while adhering to the timing parameters. \atb{2}{We expect two outcomes from the read operation. First, the simultaneously activated rows in \rl{}'s subarray store the predefined data pattern (the exact data pattern sent with the \wri{} command) because \rl{} receives the last activate command in the \apaSub{} command sequence. Second, half of the DRAM cells in the simultaneously activated rows in \rf{}'s subarray store the inverse of the predefined data pattern \emph{if \rl{} and \rf{}'s subarrays are neighbors}. This way, we determine 1) if \rl{} and \rf{} are in neighboring subarrays, and 2) how many (and which) rows \apaSub{} simultaneously activates in neighboring subarrays.}
}

\noindent\textbf{Number of \iey{1}{\rf{} and \rl{} Combinations} Tested.}~\iey{1}{We extensively perform our experiments in four randomly selected \iey{3}{neighboring} subarray pairs (i.e., a total of eight \iey{3}{subarrays}) for each bank in each tested SK Hynix module. We test every possible combination of \rf{} and \rl{} in \apaSub{} for each of the neighboring subarray pairs.\footnote{\iey{3}{We test 409,600 a total combinations of \rf{} and \rl{} row addresses for each neighboring subarray pair.}}}

\noindent\textbf{Terminology.} 
\iey{1}{To ease the understanding of \iey{2}{how many rows we can simultaneously activate in two neighboring subarrays}, we introduce a new term, \emph{\nrf{}:\nrl{} activation} where \nrf{} is the number of simultaneously activated rows in \rf{}'s subarray and \nrl{} is the number of simultaneously activated rows in \rl{}'s subarray. For example, 2:4 activation stands for simultaneously activating two rows in \rf{}'s subarray and four rows in \rl{}'s subarray.
}

\noindent\textbf{Metric.}
\iey{1}{
We define a metric called \om{2}{\emph{coverage of each unique \nrf{}:\nrl{}}} activation \om{3}{type} as the fraction of \rf{} and \rl{} combinations out of all possible combinations. For example, if we perform 2:4 activation with 100 combinations of \rf{} and \rl{} out of 1000, the coverage of 2:4 activation is 10\%.
}

\subsection{COTS DRAM Chip Characterization}
\label{subsec:char_multi_sub}
\iey{2}{This section presents our characterization of the simultaneously activating multiple rows in neighboring subarrays in SK Hynix chips. We note that we conduct experiments in DRAM modules from two other major manufacturers (i.e., Samsung and Micron). In Samsung chips, we only observe sequential two-row activation in neighboring subarrays, whereas, in Micron chips, we observe neither simultaneous nor sequential row activation in neighboring subarrays.}\footnote{\label{fn:mic-sam}{\iey{2}{We refer to \secref{sec:discussion} for a hypothesis why we observe limited capability in Micron and Samsung chips. The extended version of this paper~\cite{yuksel2024functionally} presents more results and discussion on this finding.}}}

\begin{figure}[ht]
\centering
\includegraphics[width=\linewidth]{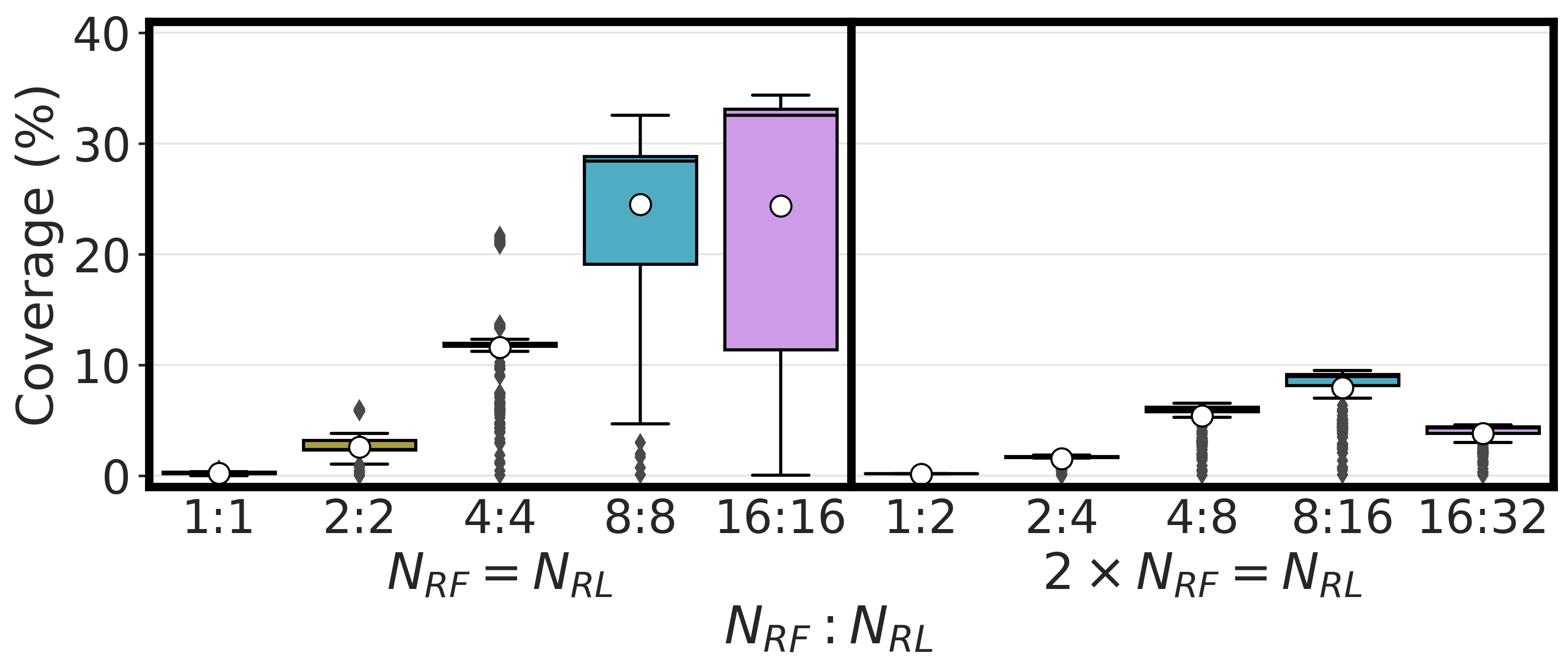}
\caption{\iey{1}{Coverage of each \nrf{}:\nrl{} activation} \om{3}{type} across tested \rf{} and \rl{} row pairs.}
\label{fig:coverage_multi_sub}
\end{figure}

\iey{1}{
\figref{fig:coverage_multi_sub} shows the distribution of observed \om{2}{coverage of each \nrf{}:\nrl{} activation} \om{3}{type} across \om{2}{all} tested combinations of \rf{} and \rl{} row addresses in a box-and-whiskers plot.\footnote{\label{fn:boxplot}{The box is lower-bounded by the first quartile (i.e., the median of the first half of the ordered set of data points) and upper-bounded by the third quartile (i.e., the median of the second half of the ordered set of data points). The interquartile range (IQR) is the distance between the first and third quartiles (i.e., box size). Whiskers show the minimum and maximum values.}} The x-axis shows the \nrf{}:\nrl{} activation type and the y-axis shows the coverage of each \nrf{}:\nrl{} activation type. We make Obsversations 1 and 2 from \figref{fig:coverage_multi_sub}.
}

\observation{\iey{1}{COTS DRAM chips can simultaneously activate multiple rows in two neighboring subarrays.}}

\iey{1}{
COTS DRAM chips can perform 1:1, 1:2, 2:2, 2:4, 4:4, 4:8, 8:8, 8:16, 16:16, and 16:32 activation with an average coverage across all \iey{2}{tested} DRAM chips of \param{0.23\%}, \param{0.15\%}, \param{2.60\%}, \param{1.53\%}, \param{11.58\%}, \param{5.42\%}, \param{24.52\%}, \param{7.95\%}, \param{24.35\%}, and \param{3.82\%}, respectively.} \iey{1}{
We observe that when we issue \apaSub{} and followed by a \wri{} command, \om{2}{all cells in} the simultaneously activated row\om{2}{(s)} in \rf{}'s subarray store the exact data pattern sent with the \wri{} command. On the other hand, half of the cells in the simultaneously activated row\om{2}{(s)} in \rl{}'s subarray stores the negated value of the data pattern sent with the \wri{} command. The remaining half retain their initial value\om{2}{s}. This observation supports our hypothesis in \secref{subsec:method_multi_sub}. 
}

\observation{COTS DRAM chips have two distinct \om{3}{sets of} \iey{1}{\nrf{}:\nrl{} activation} \om{3}{patterns} in neighboring subarrays\om{3}{: one where \nrf{}=\nrl{} and another where 2$\times$\nrf{}=\nrl{}.}} 

\iey{1}{COTS DRAM chips have two distinct \nrf{}:\nrl{} activation} \om{3}{patterns} \iey{1}{in two neighboring subarrays}: 1) N:N, where \om{2}{exactly the same number of} rows are activated \om{1}{in each} subarray \iey{1}{(i.e., \nrf{}=\nrl{}=N)} and 2) N:2N, where twice as many rows are activated \om{1}{in} \iey{1}{\rl{}'s subarray \om{1}{than} \om{2}{in} \rf{}'s (i.e., \nrf{}=N and \nrl{}=2N)}. \iey{1}{We observe that some DRAM modules have the ability to perform \om{2}{\emph{both}} N:2N and N:N activation \om{3}{patterns}, resulting in simultaneous activati\om{2}{on of} up to 48 rows \om{2}{(N=16)}. On the other hand, some DRAM modules only support the N:N activation \om{3}{pattern}, which results in simultaneous activati\om{2}{on of} up to 32 rows \om{2}{(N=16)}}.

We observe that the row addresses of \rf{} and \rl{} in the \apaSub{} command sequence determine 1) the \nrf{}:\nrl{} activation \om{3}{pattern} (i.e., N:N or N:2N) and 2) the number of simultaneously activated rows (i.e., the value of N). We hypothesize that this is due to the \om{1}{structure of the} row decoder circuitry.
\om{2}{A concurrent work~\cite{yuksel2023pulsar} describes \iey{3}{a} hypothetical row decoder design and how \apaEx{} can activate multiple rows in detail. Th\iey{3}{is hypothetical row decoder}~\cite{yuksel2023pulsar} design \om{3}{could} explain how N:N and N:2N activation patterns happen and for which addresses. Due to space limitations, we refer the reader to the concurrent work~\cite{yuksel2023pulsar} for \om{3}{more detail on the hypothetical row decoder circuitry in DRAM chips and how this row decoder design simultaneously activates multiple rows}.}
\iey{1}{From Obsvs. 1 and 2, we draw Takeaway 1.}

\takeaway{\iey{0}{COTS DRAM chips can simultaneously activate \iey{1}{multiple DRAM rows (up to 48 DRAM rows) in two neighboring subarrays.}}}

\setcounter{version}{3}
\section{NOT Operation in COTS DRAM Chips}
\label{sec:not}
\iey{2}{We demonstrate a new computation capability of commercial off-the-shelf (COTS) DRAM chips: we can perform the bitwise NOT operation in COTS DRAM chips by leveraging simultaneous multiple row activation in neighboring subarrays (\secref{sec:multi_sub}).}
\iey{1}{\atb{2}{We} simultaneously activate multiple rows in two neighboring subarrays to connect a row in one subarray (i.e., source row) to a row in a neighboring subarray (i.e., destination row) via a NOT gate in the sense amplifiers, shared by these neighboring subarrays. This connection results in \om{2}{(half of)} the destination row storing the negated value of \om{2}{(half of)} the source row (\secref{subsec:hypo_not}). We demonstrate that COTS DRAM chips can execute the NOT operation and quantitatively analyze the reliability of performing the NOT operation in \nChip{} COTS DRAM chips (\secref{subsec:char_not}).}

\subsection{NOT Operation: Key Idea}
\label{subsec:hypo_not}
\iey{1}{We leverage our key observation to enable NOT in COTS DRAM chips: simultaneous activation of multiple rows in two neighboring subarrays. Our key idea is to \iey{3}{simultaneously activate multiple rows in neighboring subarrays to connect simultaneously activated rows in neighboring subarrays through} a NOT gate in the shared sense amplifier. We use this connection to negate data in \om{2}{(half of)} one row (i.e., source row) to another row that resides in the neighboring subarray (i.e., destination row). }

\begin{figure}[ht]
\centering
\includegraphics[width=\linewidth]{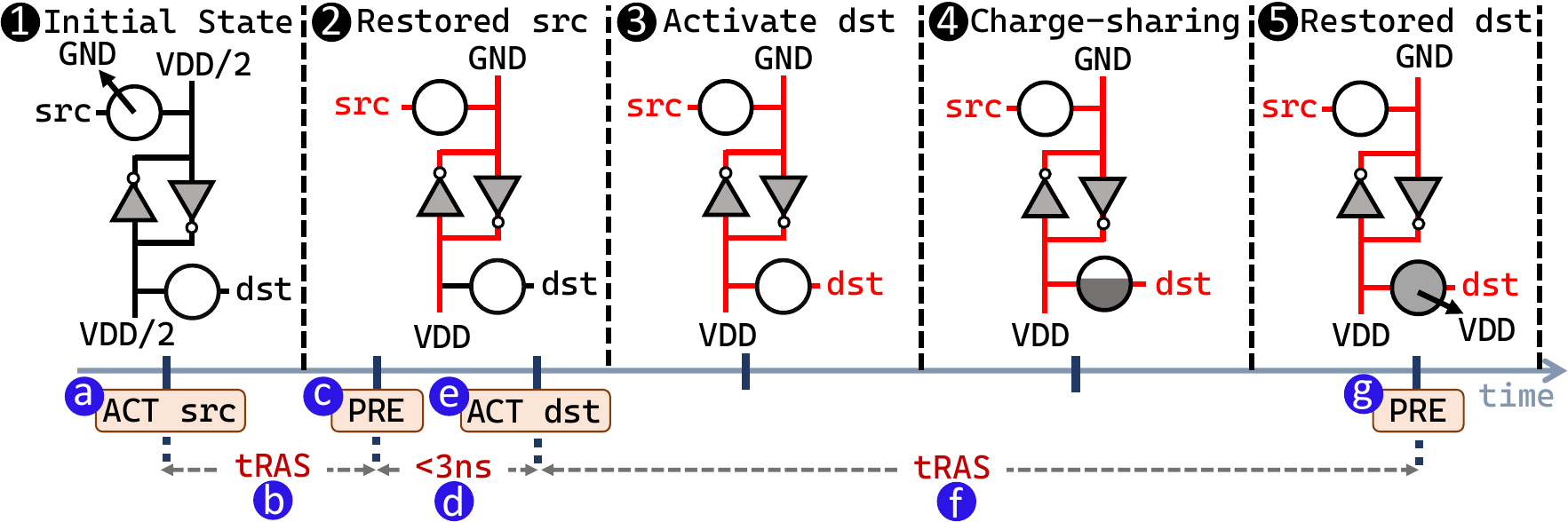}
\caption{Command sequence for performing the NOT operation in COTS DRAM chips and the state of cells during each related step.}
\label{fig:not_mech}
\end{figure}

\iey{1}{
\figref{fig:not_mech} demonstrates our key idea to enable NOT in an example of two cells (\src{} and \dst{}) \iey{1}{where \src{}'s subarray and \dst{}'s subarray are neighbors, i.e., physically adjacent.} The memory controller issues each command (shown in orange boxes below the time axis) at the corresponding tick mark, \iey{2}{and
asserted signals are highlighted in red}. Cells initially have a voltage level of ground (GND), and the bitline (i.e., \src{}'s bitline) and bitline-bar (i.e., \dst{}'s bitline) initially have a voltage level of VDD/2} (\dingOne{}).

\iey{1}{
Our key idea enables NOT operation in COTS DRAM chips in three steps. 
First, we issue an \act{} command to \src{}, i.e., \texttt{ACT}~\src{} (\dingA{}), and wait for the manufacturer-recommended \tras{} timing parameter to restore the charge of \src{} (\dingB{}). As a result, the bitline \iey{2}{reaches} the \src{} voltage (GND), whereas the bitline-bar \iey{2}{reaches} the negated \src{} voltage (VDD)} (\dingTwo{}). 
\iey{1}{
Second, we issue a \pre{} command (\dingC{}) and with violated \trp{} timing, e.g., $<$3ns (\dingD{}), we issue another \act{} command to activate \dst{}, i.e., \texttt{ACT}~\dst{} (\dingE{}). Issuing back-to-back \texttt{PRE} $\rightarrow$ \texttt{ACT}~\dst{} activates \dst{} without deactivating \src{}} (\dingThree{})\iey{2}{~\cite{gao2019computedram,gao2022frac,olgun2022pidram,olgun2021quactrng}}. \iey{1}{This results in the bitline-bar sharing its charge with \dst{} \om{2}{by} driv\om{2}{ing} the negated voltage value of \src{} (VDD) \om{2}{in}to \dst{}} (\dingFour{}).
\iey{1}{Third, we wait for the manufacturer-recommended \tras{} timing parameter (\dingF{}), which \om{2}{completely} restores the charge of \dst{}, \om{2}{and thus,} the negated value of \src{} (VDD) is written to \dst{}} (\dingFive{}). \iey{1}{Fourth, we send a \pre{} command to complete the process (\dingG{}).
}

\subsection{Experimental Methodology}
\label{subsec:method_not}

\noindent\textbf{Testing Methodology.} 
Our experiment consists of four steps. We \one{} initialize \iey{1}{the source row (\src{})'s subarray and destination row (\dst{})'s subarray with a predetermined data pattern where the subarrays of \src{} and \dst{} are neighbors, \two{} write a data pattern that is different \om{2}{from} the predetermined data pattern to \src{} }, \iey{1}{\three{} issue \apaEx{} command sequence with reduced \trp{} timing to perform the NOT operation, as described in \secref{subsec:hypo_not} and \four{} wait for \tras{} and precharge the tested bank. After the four-step procedure, we read \atb{1}{all} \om{2}{rows in \dst{}'s subarray}}\iey{3}{. If COTS DRAM chips can execute the NOT operation, half of the cells in the simultaneously activated rows in \dst{}'s subarray will store the \emph{negated} random data pattern that is written to the \src{}.\footnote{Since half of the bitlines are shared among two neighboring subarrays through a NOT gate, the \om{1}{proposed} NOT operation can negate half of the row.} We refer to simultaneously activated rows in \dst{}'s subarray as \om{2}{\emph{destination rows}.}}

\noindent\textbf{Distance Between a Row and Sense Amplifiers.} 
\iey{1}{To understand the effects of \emph{design-induced variation}~\cite{lee2017design} on the reliability of NOT operations, we analyze \atb{1}{how the} distance \atb{1}{between} activated rows (i.e., \src{} and destination rows) \atb{1}{and} the shared sense amplifiers (i.e., the sense amplifiers physically adjacent to both of the neighboring subarrays) \atb{1}{affects the reliability of NOT operations}.\footnote{\atb{1}{We are interested in analyzing the effects of design-induced variation because p}rior work~\cite{lee2017design} experimentally demonstrate\atb{1}{s} that there is \om{2}{large} design-induced variation in DRAM cells based on their physical location in the DRAM chip. \atb{1}{For example,} cells closer to the peripheral structures (e.g., sense amplifiers or wordline drivers) can be accessed faster than cells that are farther~\cite{lee2017design}.}}

\atb{1}{We leverage a widely-used reverse engineering technique~\cite{yaglikci2022hira,olgun2023experimental,orosa2021deeper,yaglikci2022understanding,kim2020revisiting,yaglikci2024svard} where we analyze RowHammer~\cite{kim2014flipping,mutlu2019retrospective,mutlu2023fundamentally} errors (i.e., bitflips) to understand how close a row is to the sense amplifiers.}~\atb{1}{The technique relies on a key characteristic of RowHammer bitflips:} when we perform single-sided RowHammer~\cite{kim2014flipping,orosa2021deeper,kim2020revisiting,yaglikci2022understanding,orosa2022spyhammer}, we observe bitflips in a row that is physically adjacent to the frequently activated (or hammered) aggressor row. If we observe bitflips in two rows, we conclude that one of them is located above the aggressor row while the other is located below the aggressor row. However, if we observe bitflips in only one row, we hypothesize that the aggressor row is physically adjacent to the sense amplifier, and thus, only one row has bitflips. By leveraging these observations, we uncover the physical order of rows in every tested subarray.

\iey{1}{We categorize the distance between activated rows and the sense amplifiers into three regions, each of which has \om{2}{one} third of all rows in the subarray: Far, Close, and Middle. ``Far'' refers to the rows that are farthest away from the sense amplifiers, ``Close'' refers to the rows that are closest to the sense amplifiers, and ``Middle'' refers to the remaining rows in the subarray.
}

\noindent\textbf{Metric.}
To quantitatively evaluate the reliability of \om{2}{the} NOT operation, we use a metric called \emph{success rate}. The success rate for a DRAM cell refers to the percentage of \iey{1}{trials \atb{1}{where the} negated value of \src{} is stored in the DRAM cell} out of all tested 10000 trials. For example, if a cell \atb{1}{(in \dst{})} stores \iey{1}{the negated value \atb{1}{(of \src{})} in 1000 trials out of 10000 trials, the success rate of the cell is 10\%. We define the \emph{average success rate} as the mean of all tested DRAM cells' success rate.} 

\noindent\textbf{The Number of Tested Instances.}
\iey{1}{We test all 16 banks in each tested SK Hynix and Samsung chip. We extensively perform our experiments in four randomly selected subarray pairs (i.e., a total of eight subarrays) in each bank. For each neighboring subarray pair, we test every possible \src{} and \dst{} row addresses \atb{1}{(e.g., 409,600 such combinations of \src{} and \dst{} row addresses for a subarray pair in an SK Hynix module)}. 
}

\noindent\textbf{Data Patterns.}
\iey{1}{We use two random data patterns (RAND1 and RAND2) and two fixed data patterns (all-1s and all-0s that fill a row with logic-1 and logic-0 values, respectively). We fill 1) the \src{} with RAND1 and 2) the rows in the \dst{}'s subarrays and the rows in the \src{}'s subarray (except \src{}) with RAND2. 
We present \atb{1}{the results for random data patterns} as we observe at most <0.1\% success rate difference between initializing rows with random data versus all-1s/0s.}

\noindent\textbf{Temperature.} We perform our experiments \om{2}{at} five temperature levels: 50$^{\circ}$C, 60$^{\circ}$C, 70$^{\circ}$C, 80$^{\circ}$C, and 95$^{\circ}$C. \iey{1}{All experiments are conducted at 50$^{\circ}$C unless stated otherwise.}

\subsection{COTS DRAM Chip Characterization}
\label{subsec:char_not}
\iey{1}{This section presents our rigorous characterization of the reliability of NOT operations in SK Hynix and Samsung chips. While we test all three major manufacturers (i.e., SK Hynix, Samsung, and Micron), we note that 1) NOT operations have one destination row in Samsung chips, as we only observe sequential two-row activation in neighboring subarrays (\secref{subsec:char_multi_sub}) in Samsung chips, and 2) we do \emph{not} observe NOT operations in Micron chips.}\iey{2}{\footref{fn:mic-sam}}

\begin{figure}[ht]
\centering
\includegraphics[width=0.77\linewidth]{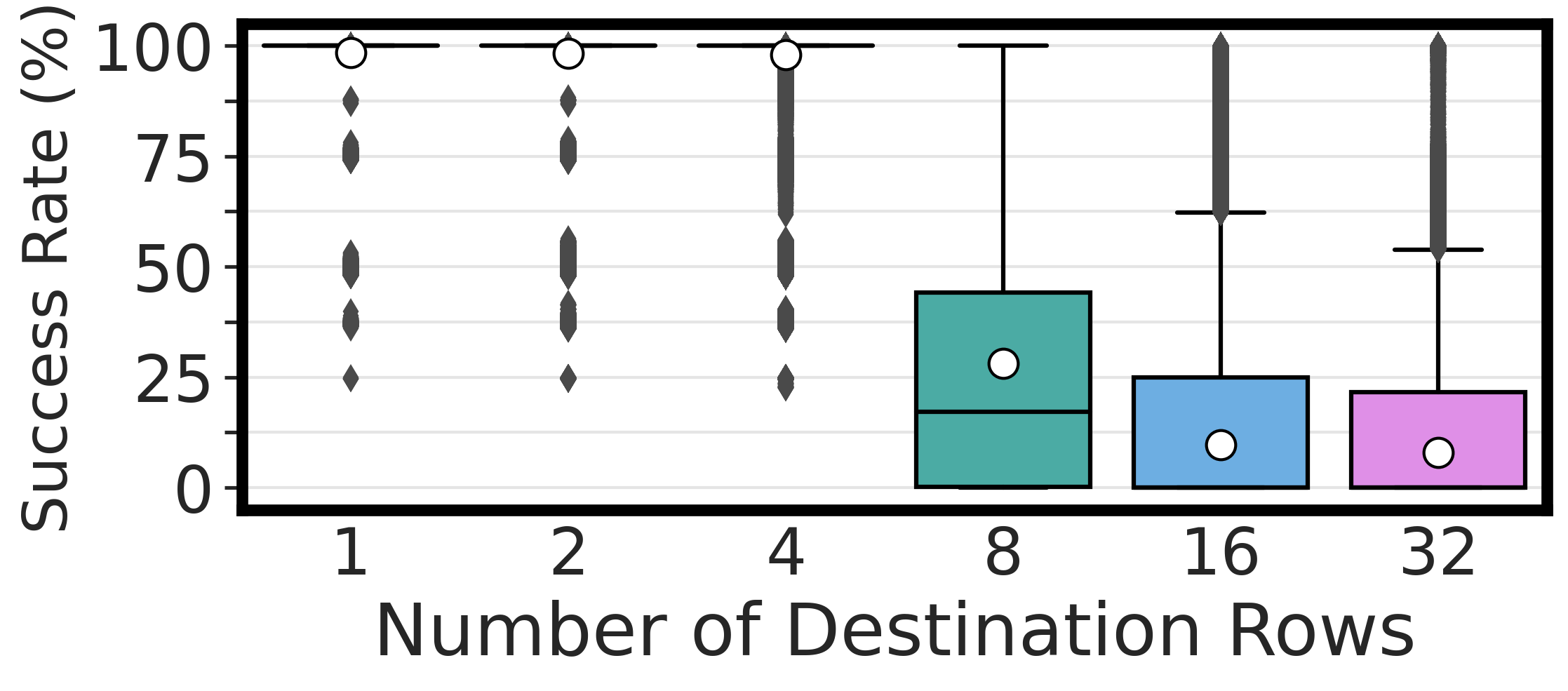}
\caption{\om{1}{Success rate} of the NOT operation in COTS DRAM chips with varying numbers of destination rows.}
\label{fig:not_1}
\end{figure}

\noindent\textbf{Number of \om{3}{D}estination \om{3}{R}ows.} We analyze the reliability of NOT operation\atb{1}{s} with \param{1, 2, 4, 8, 16, and 32} destination rows. \figref{fig:not_1} shows the \iey{0}{success rate} distribution across DRAM cells \atb{1}{in all tested \iey{3}{SK Hynix and Samsung} DRAM chips} in a box and whiskers plot as we vary the number of destination rows from \param{1} to \param{32}.\footref{fn:boxplot} \iey{2}{We make \om{3}{Observations} 3 and 4 from \figref{fig:not_1}.}

\observation{\iey{1}{In every tested number of destination rows, there is at least one \om{2}{DRAM cell} where we can perform the NOT operation with a 100\% success rate.}}

\iey{1}{COTS DRAM chips can perform NOT operation\atb{1}{s} with 1, 2, 4, 8, 16, and 32 destination rows. We observe that in every number of destination rows tested, there \atb{1}{is at least one} \iey{2}{cell}
that we can use to perform the NOT operation with 100\% \iey{0}{success rate}}.

\observation{As the number of destination rows increases, more DRAM cells \iey{0}{produce incorrect results, leading to a decrease in success rate.}}

For example, the average \iey{0}{success rate} of \om{1}{the} NOT operation with one destination row is \param{98.37}\%, \om{1}{but it is only} 7.95\% with 32 destination rows. \atb{1}{We hypothesize that the decrease in success rate could be due to the increase in total bitline capacitance that a sense amplifier must drive as the number of simultaneously activated rows increases. A sense amplifier has to drive the negated voltage level of a source row's cell into} 32 destination cells when we perform the NOT operation with 32 destination rows, whereas the sense amplifier needs to drive the negated voltage value into \emph{only} one destination cell when we perform the NOT operation with a \emph{single} destination row. \atb{1}{Weaker sense amplifiers could incorrectly drive the negated value of a source row's cell into multiple destination rows, \om{2}{and DRAM sense amplifiers are clearly \om{3}{\emph{not}} designed for this purpose}.}

\begin{figure}[ht]
\centering
\includegraphics[width=0.8\linewidth]{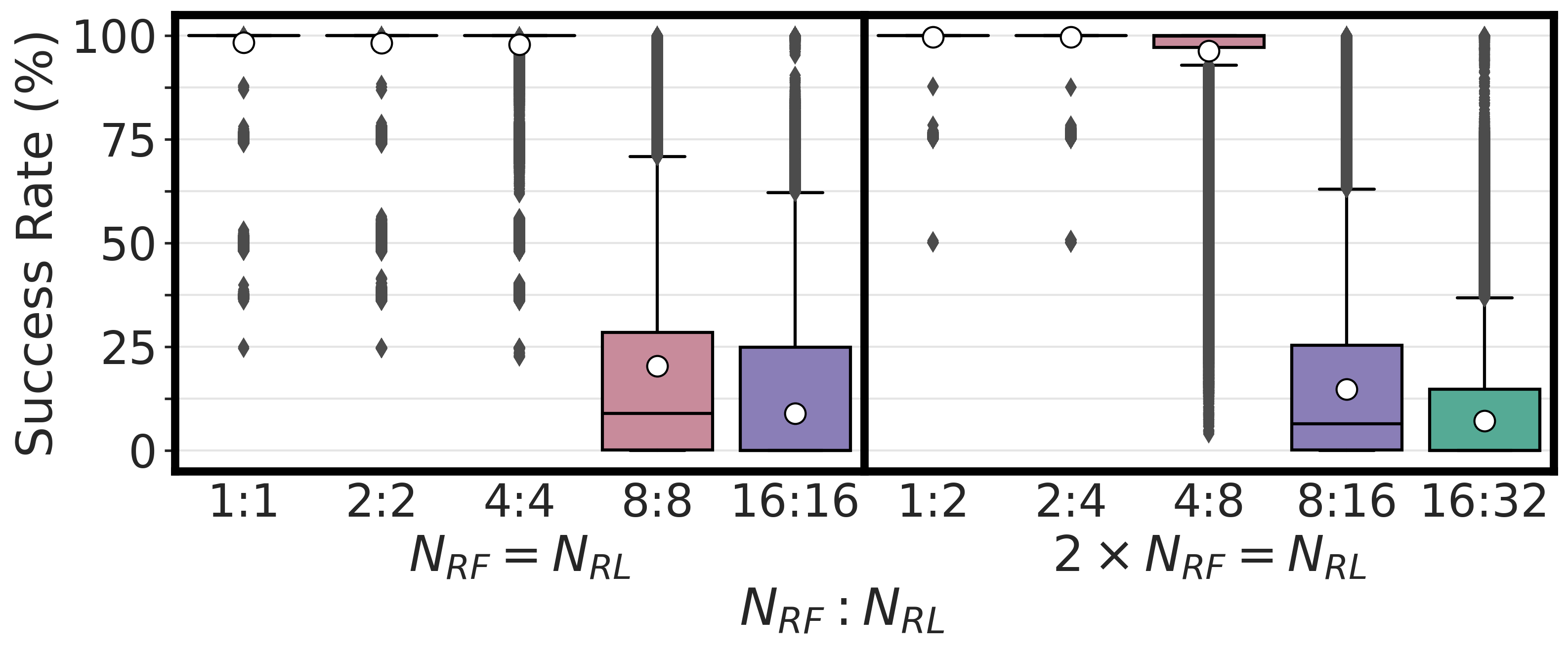}
\caption{\iey{1}{Success rate of the NOT operation \om{3}{vs.} \nrf{}:\nrl{} activation \om{3}{type}.}}
\label{fig:not_2}
\end{figure}

\noindent\textbf{\iey{3}{Patterns of Multiple Row Activation in Neighboring Subarrays.}} \iey{1}{We analyze the effect of \nrf{}:\nrl{} activation (where \nrf{} is the number of the simultaneously activated rows in \src{}'s subarray, and \nrl{} is the number of simultaneously activated rows in \dst{}'s subarray), on the reliability of the NOT operation. In \secref{subsec:char_multi_sub}, we observe two distinct \iey{3}{sets of \nrf{}:\nrl{} activation patterns}: 1) N:N where \nrf{}=\nrl{} and 2) N:2N where 2$\times$\nrf{}=\nrl{}}.~\figref{fig:not_2} shows the \iey{1}{distribution of} \iey{0}{success rate across DRAM cells} \iey{1}{in a box and whiskers plot}, with the x-axis representing \nrf{}:\nrl{} \iey{3}{activation types} \iey{0}{(e.g., 1:2 represents one row is activated in \src{}'s subarray and two rows are activated in \dst{}'s subarray).} \iey{2}{We make \om{3}{Observation} 5 from \figref{fig:not_2}.}

\observation{\iey{3}{The} N:2N activation \om{3}{pattern} \atb{1}{results in} slightly higher \iey{0}{success rate} than \iey{3}{the} N:N activation \om{3}{pattern}.}

Using \om{3}{the} N:2N \iey{1}{activation \om{3}{pattern}} to perform the NOT operation exhibits \param{9.41}\% higher \iey{0}{average success rate} than using \om{3}{the} N:N activation \om{3}{pattern} to perform the NOT operation. \iey{1}{We hypothesize that this is due to the total number of activated rows in two neighboring subarrays. For example, to perform the NOT operation with 16 destination rows,} \iey{2}{sense amplifiers have to simultaneously drive 32 (24) rows when using \om{3}{the} N:N (N:2N) \om{3}{pattern}. This results in sense amplifiers driving more rows in \om{3}{the} N:N \om{3}{pattern} than \om{3}{the} N:2N \om{3}{pattern}. Due to process variation, not all sense amplifiers are strong enough to correctly drive the \om{3}{corresponding voltage value}  into multiple rows (i.e., \src{}'s voltage value into the activated rows in \src{}'s subarray and the negated voltage value of \src{} into destination rows) at once. This leads to a decrease in the success rate of the NOT operation as the number of simultaneously activated rows increases.}

\begin{figure}[ht]
\centering
\includegraphics[width=0.62\linewidth]{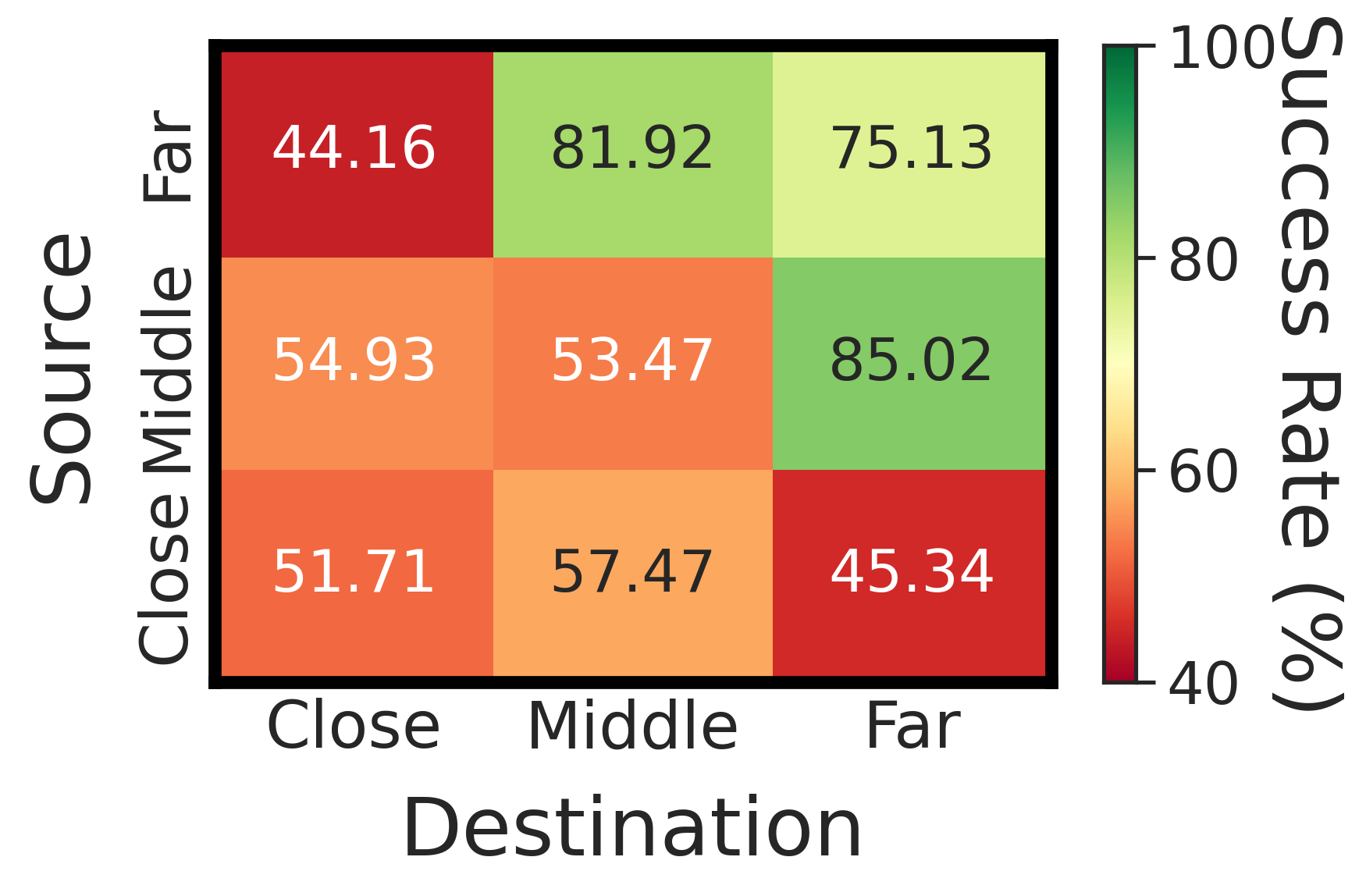}
\caption{S\iey{0}{uccess rate} of \iey{2}{the} NOT operation based on the distance of activated rows to sense amplifiers.}
\label{fig:not_3}
\end{figure}

\noindent\textbf{\iey{2}{Distance} to the \om{3}{S}ense \om{3}{A}mplifier.} 
\figref{fig:not_3} shows the average \iey{0}{success rate} of \iey{1}{the NOT operation (i.e., the mean success rate of every tested DRAM cell in all tested destination rows)} using a heatmap plot based on the proximity \atb{1}{of \src{} (y-axis) and all activated destination rows (x-axis) to the sense amplifiers that physically reside between (i.e., shared by) \src{} and \dst{}'s subarrays.}\iey{2}{We make \om{3}{Observation} 6 from \figref{fig:not_3}.}

\observation{\iey{1}{The success rate} of the NOT operation significantly varies based on the proximity \atb{1}{of the} activated rows (\src{} and the destination rows) to the sense amplifier\atb{1}{s}.}

\iey{1}{For example, when the source row is in the middle of the subarray, and the destination rows are far from the sense amplifier (i.e., Middle-Far), the average \iey{0}{success rate} is \param{85.02}\%. However, when the source row is far from the sense amplifiers and the destination rows are close to the sense amplifiers (i.e., Far-Close), the average success rate is 44.16\%.}~\iey{2}{We hypothesize that \emph{design-induced variation}~\cite{lee2017design} \atb{2}{could explain} the large variation in the success rate. \atb{2}{Design-induced variation} causes cell access latency characteristics to vary deterministically depending on the physical location of cells in the memory chip, including \atb{2}{the cell's} distance to the sense amplifier.}

\begin{figure}[ht]
\centering
\includegraphics[width=\linewidth]{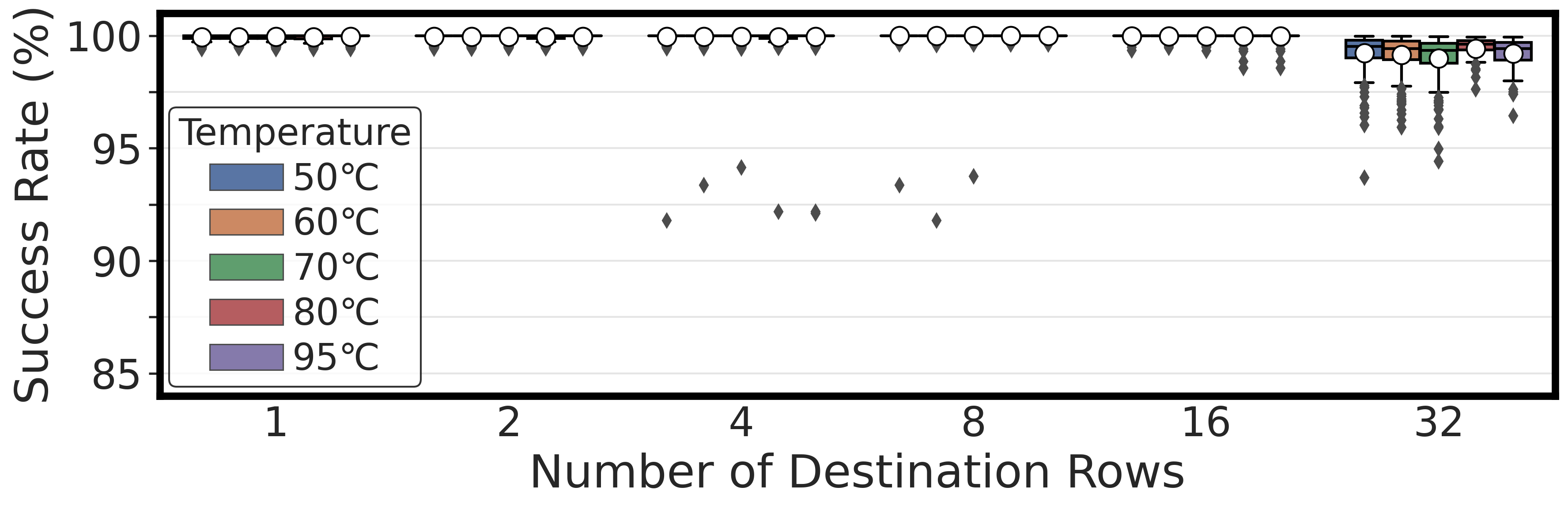}
\caption{\atb{1}{Success rate} of \iey{2}{the} NOT operation \atb{1}{at different DRAM chip temperatures}.}
\label{fig:not_4}
\end{figure}

\noindent\textbf{Temperature.} In this experiment, we use destination \iey{2}{cells} that can perform NOT operations with $>$90\% \iey{0}{success rate} at 50$^{\circ}$C.\footnote{\label{fn:testing}{\iey{2}{To maintain a reasonable testing time, w}e test destination \iey{2}{cells} with $>$90\% \iey{0}{success rate} from every tested destination row, obtained from all tested SK Hynix DRAM modules \iey{2}{(on average $\approx$1.39 billion \atb{2}{such} cells per module).}}}
\figref{fig:not_4} shows the \iey{0}{success rate} distribution across \atb{1}{all tested} DRAM cells \atb{1}{for} five different temperature levels: 50$^{\circ}$C, 60$^{\circ}$C, 70$^{\circ}$C, 80$^{\circ}$C, and 95$^{\circ}$C. \atb{1}{The figure clusters boxes into six groups based on the number of activated destination rows (x-axis). In each cluster \om{2}{of boxes, temperature increases} from left to right.} \iey{2}{We make \om{3}{Observation} 7 from \figref{fig:not_4}.} 

\observation{\iey{0}{Temperature has a \atb{1}{small} effect on the reliability of NOT operations in \atb{1}{COTS} DRAM chips.}}

For example, performing the NOT operation with \param{32} destination rows (which is the most vulnerable operation \atb{1}{to temperature variation}) \atb{1}{exhibits} a \param{0.20}\% variation in average \iey{0}{success rate} when the temperature \atb{1}{increases} \atb{1}{from 50$^{\circ}$C} \om{2}{to 95$^{\circ}$C}.

\takeaway{COTS DRAM chips can perform NOT operation\atb{1}{s with} up to 32 destination rows. \atb{1}{NOT operations are} highly resilient to temperature changes.}

\begin{figure}[ht]
\centering
\includegraphics[width=\linewidth]{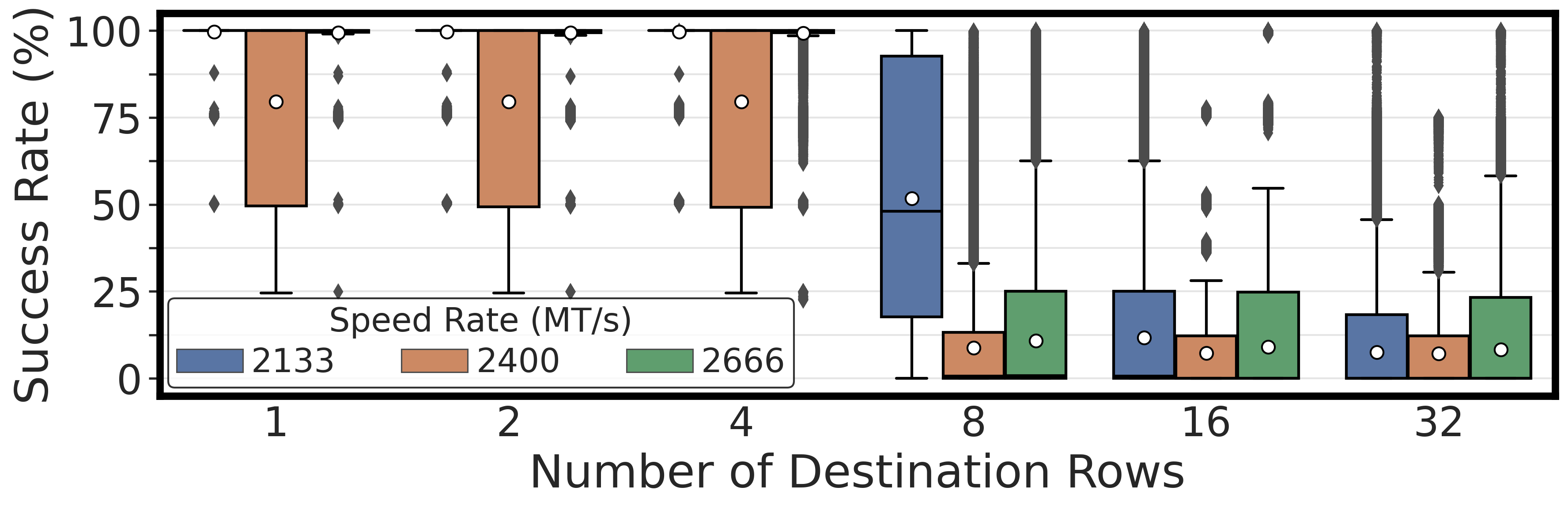}
\caption{Success rate of \iey{2}{the} NOT operation for different DRAM speed rate\om{2}{s}.}
\label{fig:speed_not}
\end{figure}

\noindent\textbf{DRAM Speed Rate.}
We analyze the effect of DRAM speed rate on the \iey{0}{success rate} of NOT operations in SK Hynix DRAM modules.~\figref{fig:speed_not} shows the \iey{0}{success rate} distribution across DRAM cells, with the x-axis representing the number of destination rows and the hue showing the DRAM speed rate. \iey{2}{In each cluster of boxes, the DRAM speed rate increases from left to right.} \iey{2}{We make \om{3}{Observation} 8 from \figref{fig:speed_not}.}

\observation{\iey{2}{The \iey{0}{success rate} of a NOT operation significantly varies across tested DRAM speed rates.}}

For example, for the NOT operation with 4 destination rows, \iey{2}{average \iey{0}{success rate} decreases by \param{20.06\%}} as \om{3}{DRAM} speed \om{3}{rate} increases from 2133 MT/s to 2400 MT/s. However, \iey{3}{the average success rate of the NOT operation with 4 destination rows} \iey{2}{increases by \param{19.76\%}} from 2400 MT/s to 2666 MT/s.

\begin{figure}[ht]
\centering
\includegraphics[width=\linewidth]{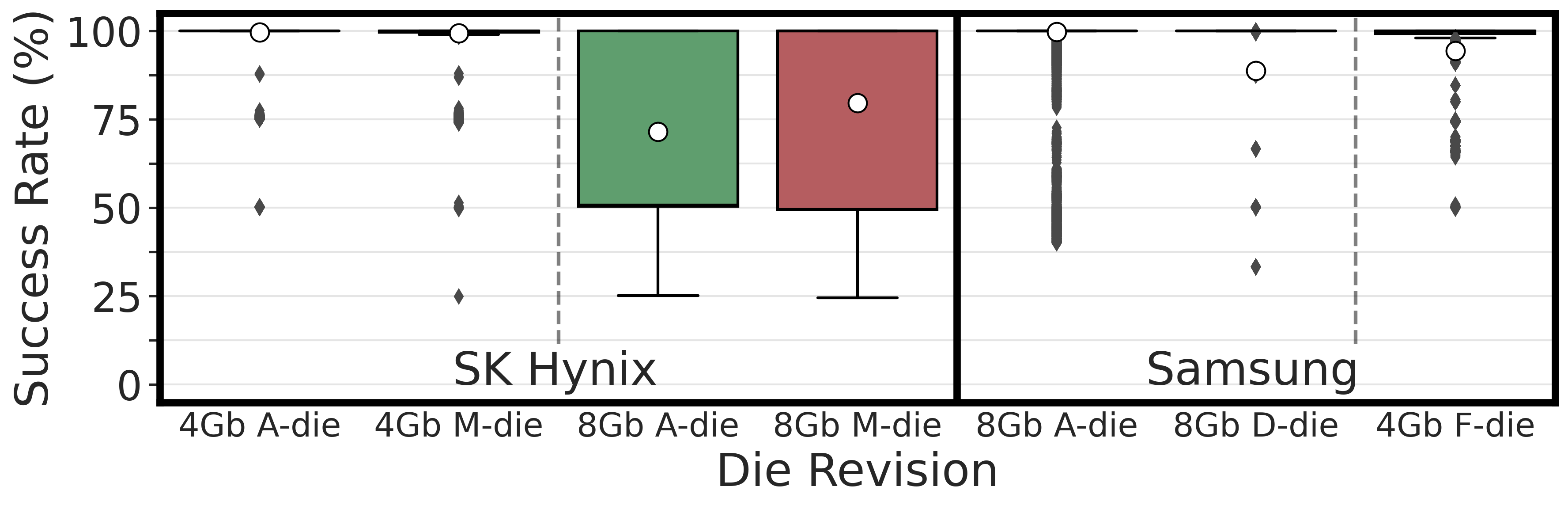}
\caption{Success rate of \iey{2}{the} NOT operation for different \om{3}{chip density and die revision combinations} for two major manufacturers.}
\label{fig:die_not}
\end{figure}

\noindent\textbf{\iey{3}{Chip Density and Die Revision}.}
\iey{0}{We analyze \iey{3}{the effect of chip density and die revision on} the reliability of \atb{1}{NOT operations}. \atb{1}{To do so, we use} DRAM modules from two major manufacturers spanning different} \om{3}{chip densities and die revisions}.\footref{fn:die} \atb{1}{For this analysis, w}e \atb{1}{show the results for} perform\atb{1}{ing} NOT operations with \atb{1}{only} one destination row.\footnote{\iey{2}{W\atb{2}{e use one destination row} as Samsung chips cannot perform the NOT operation with more \atb{2}{than one} destination row. We show more comprehensive results for more \atb{2}{than one} destination row \atb{2}{using SK Hynix chips\om{3}{.}} in the extended version of this paper~\cite{yuksel2024functionally}}} \figref{fig:die_not} shows the \iey{0}{success rate} distribution across DRAM cells, with the x-axis representing the module's \om{3}{chip density, die revision} and manufacturer. \iey{2}{We make \om{3}{Observation} 9 from \figref{fig:die_not}.}

\observation{\iey{3}{Chip density and die revision affect} the \iey{0}{success rate} of NOT operations \atb{1}{for} each tested manufacturer.}

\iey{1}{For example, in SK Hynix DRAM chips,} the \atb{1}{average} \iey{0}{success rate} decreases by 8.05\% from 8Gb M-die to 8Gb A-die. In Samsung DRAM chips, the \atb{1}{average} \iey{0}{success rate} decreases by 11.02\% from A-die to \om{3}{D}-die.

\takeaway{NOT operation's reliability significantly varies across DRAM \om{3}{chips with different} \iey{2}{speed rate\iey{3}{s},} die revisions\iey{2}{, and chip densities}.}

\setcounter{version}{3}

\section{\atb{1}{Many-Input} NAND, NOR, AND, and OR in COTS DRAM Chips}
\label{sec:nand}
\iey{1}{We demonstrate a new computational capability of commercial off-the-shelf (COTS) DRAM chips: we can perform many-input (i.e., up to 16-input) NAND, NOR, AND, and OR operations in COTS DRAM chips. \om{3}{First,} \atb{2}{w}e leverage simultaneous multiple-row activation in neighboring subarrays \om{2}{(\secref{sec:multi_sub})} to manipulate the bitline voltage in one of the neighboring subarrays, which we refer to as the \emph{reference subarray}. \atb{2}{Manipulating the bitline voltage in the reference subarray} enables us to perform many-input AND and OR operations in the neighboring subarray of \atb{2}{the} reference subarray, which we refer to as the \emph{compute subarray}. Second, we leverage the NOT gate connection \atb{2}{(through sense amplifiers)} between simultaneously activated rows in neighboring subarrays to realize NAND and NOR operations, such that \atb{2}{the result of an} AND (OR) operation \atb{2}{on a set of input operands} in the compute subarray \atb{2}{is inverted to become the result of a} NAND (NOR) operation \atb{2}{on the same set of input operands} in the reference subarray. This section describes our key idea to enable NAND, NOR, AND, and OR operations in detail (\secref{subsec:hypo_nand}), our experimental methodology (\secref{subsec:method_nand}), and presents our rigorous characterization of if and how well COTS DRAM chips \om{2}{can} perform these Boolean operations (\secref{subsec:char_nand}).}

\subsection{\atb{1}{Many-Input Boolean Operations: Key Idea}}
\label{subsec:hypo_nand}

\atb{1}{Our key idea is to leverage an implication of simultaneous \atb{2}{multiple row activation} in neighboring subarrays for sense amplifier operation (see \secref{subsec:dram_org} for a\om{2}{n} overview of how a sense amplifier operates): the sense amplifier terminals exhibit voltage levels equivalent to \emph{a function of the values stored in many simultaneously activated cells}.}

\om{2}{To provide an intuitive} \atb{1}{high-level explanation of our key idea, we consider that a sense amplifier works in two steps where 1) it compares the voltage levels on its two terminals, \emph{terminal A} and \emph{terminal B}, and 2) drives terminal A with the result of the comparison (i.e., if A>B, the result is a high voltage value and vice versa) and terminal B with the inverse of the result. When this sense amplifier operates under standard operating conditions \om{2}{(i.e., single row activation)}, the result of the comparison is nothing but the value stored in the activated cell. This is because terminals A and B are initially precharged (i.e., \om{2}{they} store VDD/2), and the activation of a cell perturbs the bitline voltage (i.e., one of the two terminals stores a value lower or higher than VDD/2 when the sense amplifier compares their values) towards the value stored in the activated cell. However, when we activate multiple cells connected to terminals A and B, we can make both terminal A and terminal B exhibit a wider variety of voltage levels when the sense amplifier compares the two terminals' voltage levels. The wide variety of voltage levels is essentially a function of \atb{2}{the} values stored in the simultaneously activated cells. Next, we explain in detail how \atb{2}{this function of the values stored in the cells} could be combined with the sense amplifier's fundamental comparator operation to implement many-input Boolean AND, OR, NAND, and NOR operations.}

\subsubsection{\iey{1}{Performing \atb{2}{Two-}Input AND and OR}}
\label{subsubsec:and_or}
\begin{figure}[ht]
\centering
\includegraphics[width=\linewidth]{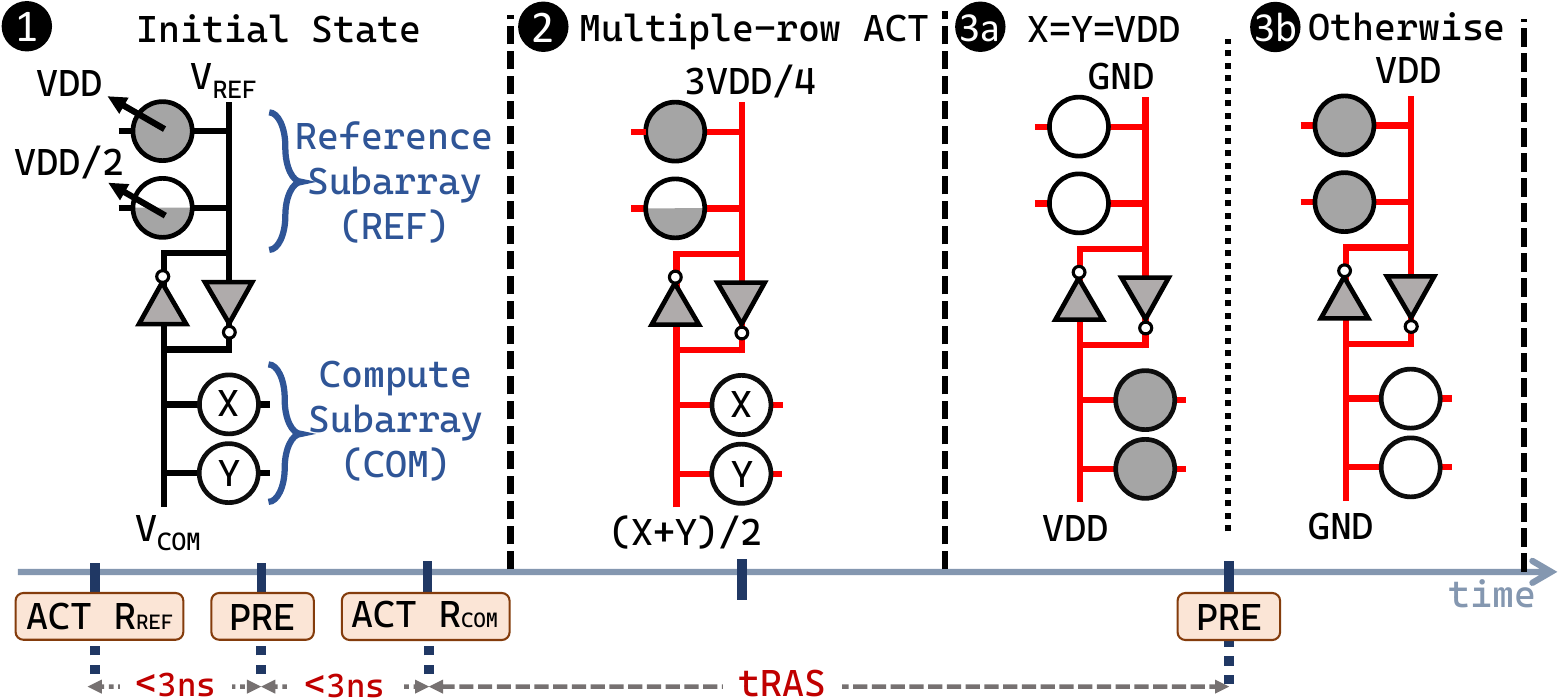}
\caption{Performing two-\om{2}{input} AND and NAND operations in COTS DRAM chips.}
\label{fig:and_mech}
\end{figure}

\iey{1}{
\figref{fig:and_mech} illustrates an example of an in-DRAM two-\om{2}{input} AND operation. The memory controller issues each command (shown in orange boxes below the time axis) at the corresponding tick mark\iey{2}{, and asserted signals are highlighted in red}. In this figure, we have two neighboring subarrays: the \emph{reference subarray} and the \emph{compute subarray}, each containing two cells. \vref{} is the voltage level of the reference subarray's bitline, and \vcom{} is the voltage level of the compute subarray's bitline.\footnote{\label{footnote:model}\atb{1}{\om{2}{To simplify the explanation,} we assume that the bitline has no capacitance (i.e., after charge sharing, the bitline's voltage is the \emph{mean} voltage value stored in DRAM cells that contribute to charge sharing) and the modeled \iey{2}{DRAM} circuitry is not subject to the effects of process variation. We later discuss and experimentally demonstrate (\secref{subsec:char_nand}) how process variation affects the operation described}.} Assume \om{2}{that the} \apaAnd{} command sequence with reduced timing simultaneously activates all four rows in these two subarrays where \rref{} points to a row in the reference subarray and \rcom{} points to a row in the compute subarray. Initially, \atb{2}{1) we store VDD in one cell and VDD/2 in the other cell \atb{2}{in the reference subarray}, and 2) we store a voltage level of X in one cell and Y in the other cell in the compute subarray} (\dingOne{}).\footnote{We store VDD/2 in a DRAM cell using an operation called Frac~\cite{gao2022frac} \atb{2}{in COTS DRAM chips}.} 
}

\iey{1}{
 To perform a two-\om{2}{input} AND operation, we first issue \atb{2}{one} \apaAnd{} command sequence (\dingOne{}). Doing so activates four rows simultaneously and enables charge-sharing between their bitlines. At the end of charge-sharing, \vref{} becomes 3VDD/4 (i.e., the mean of VDD and VDD/2) and \vcom{} becomes (X+Y)/2 (\dingTwo{}).\footref{footnote:model} The sense amplifier then kicks in and amplifies the voltage difference between \vref{} and \vcom{}. If X and Y have VDD (i.e., \vcom{}=VDD), \vcom{} is higher than \vref{}, resulting in VDD in the compute subarray's activated cells and GND in the reference subarray's activated cells (\circledt{black}{3a}). Otherwise, \vcom{} \om{2}{is} lower than \vref{}, resulting in GND in the compute subarray's activated cells and VDD in the reference subarray's activated cells (\circledt{black}{3b}). After waiting for \tras{}, we issue a \pre{} command to complete the two-input AND operation.
}

\iey{1}{
We can perform a two-\om{2}{input} OR operation when we store GND and VDD/2 in the activated cells of the reference subarray. This results in \vref{} becoming VDD/4 in \figref{fig:and_mech}-\dingTwo{}. \om{2}{T}hus\om{3}{,} the output will be \om{2}{i)} logic-1\om{2}{,} if at least one activated cell of the compute subarray stores VDD \om{2}{ii)} logic-0\om{2}{, otherwise}.
}

\subsubsection{\iey{2}{\atb{2}{Performing} N-Input AND and OR}}
\label{subsubsec:n_input}
\iey{2}{
We \atb{2}{perform} $N$-input AND and OR operations by activating $N$ rows in the reference subarray and $N$ rows in the compute subarray. \atb{2}{To simplify our explanations}, we refer to \vref{} as \vand{} when we execute the AND operation and as \vor{} when we execute the OR operation.
}

\iey{2}{
For an $N$-input AND operation, the output has to be i) logic-1, if all $N$ rows in the compute subarray store $VDD$, ii) logic-0, otherwise. This is because the AND operation outputs logic-1 \atb{2}{only} when all \atb{2}{of its} inputs are logic-1. \atb{2}{That is, the AND operation should output logic-1 only when \vcom{} is VDD, and output logic-0 for all other possible \vcom{} values.} Thus, \vand{} \atb{2}{(the bitline voltage in the reference subarray)} has to be inbetween $(N-1)*$VDD$/N$ \atb{2}{(the highest \vcom{} value for the AND operation to output logic-0)} and VDD \atb{2}{(the lowest and the only \vcom{} value for the AND operation to output logic-1).}
Consequently, the output will be i) logic-1 when all inputs are $VDD$ (i.e., \vcom{}=VDD and thus \vcom{}$>$\vand{}) and ii) logic-0, when at least one of the inputs store $GND$ (i.e., \vcom{}$\leq$$(N-1)*$VDD$/N$ and thus \vcom{}$<$\vand{}). 
}

\iey{2}{
For an $N$-input OR operation, the output has to be i) logic-0, if all $N$ rows in the compute subarray store $GND$, ii) logic-1, otherwise. This is because the OR operation outputs logic-0 only when all of its inputs are logic-0. \atb{2}{That is, the OR operation should output logic-0 only when \vcom{} is GND, and output logic-1 for all other possible \vcom{} values.} Thus, \vor{} (the bitline voltage in the reference subarray)  has to be inbetween GND (the highest and the only \vcom{} value for the OR operation to output logic-0) and VDD$/N$ (the lowest \vcom{} value for the OR operation to output logic-1). As a result, the output will be i) logic-0 when all inputs are $GND$ (i.e., \vcom{}=GND and thus \vcom{}$<$\vor{}) and ii) logic-1, when at least one of the inputs store $VDD$ (i.e., \vcom{}$\geq$VDD$/N$ and thus \vcom{}$>$\vor{}). 
}

\begin{figure}[ht]
\centering
\includegraphics[width=\linewidth]{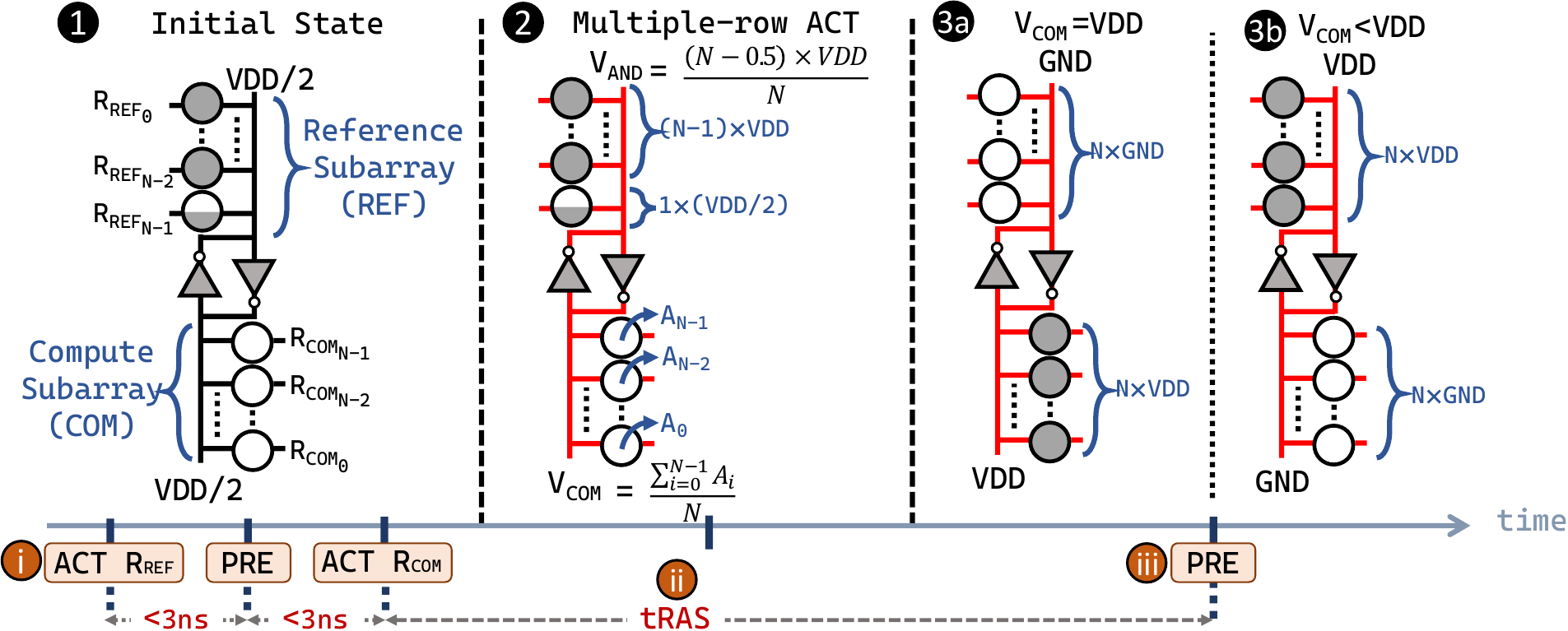}
\caption{\iey{3}{Performing N-input AND and NAND operations in COTS DRAM chips.}}
\label{fig:n_input_and}
\end{figure}

\noindent\textbf{\iey{1}{Key Mechanism.}}
\iey{1}{
\iey{3}{\figref{fig:n_input_and} illustrates our key mechanism to perform an N-input AND operation in COTS DRAM chips.} To implement \om{2}{an} $N$-input AND (OR) operation, our key mechanism consists of \param{three} steps\om{2}{.} First, for \om{2}{the} AND (OR) operation, we store VDD (GND) in $N-1$ \atb{2}{of the simultaneously activated rows in the} reference subarray \iey{3}{($\{$\rrefx{0},..,\rrefx{N-2}$\}$)}, and VDD/2 in the \atb{2}{other simultaneously activated row} in the reference subarray \iey{3}{(\rrefx{N-1})}. \iey{2}{By doing so, for the AND operation, \atb{2}{we set} \vand{} \atb{2}{to} $(N-0.5)*$VDD$/N$ \iey{3}{(}\dingTwo{}\iey{3}{)} whereas for the OR operation, \atb{2}{we set} \vor{} \atb{2}{to} $0.5*$VDD$/N$.}
Second, we issue \apaAnd{} with reduced timings to simultaneously activate $N$ rows in the reference subarray \iey{3}{($\{$\rrefx{0},..,\rrefx{N-1}$\}$)} and $N$ rows in the compute subarray \iey{3}{($\{$\rcomx{0},..,\rcomx{N-1}$\}$)}, \om{2}{as described in \secref{subsubsec:and_or} \iey{3}{(}\circledt{bronze}{i}\iey{3}{)}}. Third, we wait for the manufacturer-recommended \tras{} timing parameter \iey{3}{(}\circledt{bronze}{ii}\iey{3}{)}, which overwrites the activated cells in the compute subarray with the result of the AND (OR) operation \iey{3}{(}\circledt{black}{3a} and \circledt{black}{3b}\iey{3}{)}. Fourth, we issue a \pre{} command to complete the operation \iey{3}{(}\circledt{bronze}{iii}\iey{3}{)}.
}

\subsubsection{\iey{1}{Performing Many-Input NAND and NOR}}
\label{subsubsec:nand_nor}
\iey{1}{
\atb{2}{We} leverage the NOT operation (\secref{sec:not}) and many-input AND and OR operations \om{2}{(\secref{subsubsec:n_input})} \atb{2}{to implement NAND and NOR} as \atb{2}{a} NAND (NOR) operation is simply \atb{2}{a} chain of NOT$\rightarrow$AND (NOT$\rightarrow$OR) operations.
During the AND (OR) operation, \vcom{} becomes the output of the AND (OR) operation, and, at the same time, \vref{} becomes the output of the NAND (NOR) operation. For example, in \figref{fig:n_input_and}, where \iey{3}{we demonstrate how we perform an N-input} AND operation, \vref{} \om{2}{has} the negated value of \vcom{} \om{2}{(\circledt{black}{3a} and \circledt{black}{3b})}, resulting in the activated cells in the reference subarray \atb{2}{to store} the output of an N-input NAND operation.
}

\subsection{Experimental Methodology}
\label{subsec:method_nand}
\iey{1}{
The goal of our experiment is to understand if and how well COTS DRAM chips can perform NAND, NOR, AND, and OR operations. To perform these operations, our key idea relies on simultaneously activating an equal number of rows in neighboring subarrays (i.e., \om{2}{the} N:N activation \om{2}{pattern}) as described in \secref{subsec:hypo_nand}. We refer to the DRAM command sequence that activates N rows in each neighboring subarray as \apaAnd{} where \rref{} points to a row in the reference subarray and \rcom{} points to a row in the compute subarray.
}

\noindent\textbf{\iey{1}{Testing Methodology.}} 
\iey{1}{
Our experiment consists of four steps. First, we initialize N rows in the reference subarray. For the AND/NAND (OR/NOR) operation, N-1 rows are initialized with logic-1 (logic-0). We store VDD/2 in the remaining row by performing the Frac operation~\cite{gao2022frac}. Second, we store \atb{2}{N} input operands in N rows in the compute subarray. Third, we issue \om{2}{the} \apaAnd{} command sequence with reduced \tras{} and \trp{} timings. Fourth, we wait for manufacturer-recommended \tras{} timing and precharge the tested bank. After the four-step procedure, we read all rows in \atb{2}{the reference and the compute subarrays}.}
If \atb{2}{a} COTS DRAM chip \atb{2}{can perform} AND/NAND (OR/NOR) operations, \atb{2}{the} simultaneously activated N rows in the compute subarray will \atb{2}{store} the result of the AND (OR) operation, and the simultaneously activated N rows in the reference subarray will \atb{2}{store} the result of the NAND (NOR) operation.

\noindent\textbf{\iey{1}{Metric.}}
\iey{1}{
We use the same metric \atb{2}{that} we use to evaluate the reliability of \atb{2}{a} NOT operation: \emph{success rate}. The success rate for a DRAM cell refers to the percentage of trials where the correct output of \atb{2}{a} Boolean operation (i.e., AND or OR operation for a cell in the compute subarray and NAND or NOR operations for a cell in the reference subarray) is stored in \atb{2}{the} DRAM cell out of all tested 10000 trials. For example, if a cell in the compute subarray stores the correct output of the 16-input AND operation in 1000 trials out of 10000 16-bit AND operation trials, the success rate of the cell is 10\% for the 16-input AND operation. We define the \emph{average success rate} as the mean of all tested DRAM cells' success rate.
}

\noindent\textbf{\iey{1}{\iey{2}{N}umber of Tested DRAM Cells.}}
\iey{1}{We test all 16 banks in each tested SK Hynix chip. We extensively perform our experiments in four randomly selected neighboring subarray pairs (i.e.,
a total of eight subarrays) in each bank. For each neighboring subarray pair, we test DRAM cells that can perform NOT operations with $>$90\% success rate at 50$^{\circ}$C.\footref{fn:testing}
}

\noindent\textbf{\iey{1}{Data Pattern.}}
\iey{1}{We use two data patterns: 1) the all-1s/0s patterns, where \atb{2}{each row in the compute subarray is filled completely either with logic-1 or logic-0 values (e.g., for 4-input operations there are 16 unique such data patterns because each of the simultaneously activated four rows can either store all-1s or all-0s),} 
and 2) the random data pattern, where we fill \atb{2}{each row in the compute subarray with random data}. All experiments are conducted using the random data pattern unless stated otherwise.
}

\noindent\textbf{\iey{1}{Temperature.}}
\iey{1}{We perform our experiments at five temperature levels: 50$^{\circ}$C, 60$^{\circ}$C, 70$^{\circ}$C, 80$^{\circ}$C, and 95$^{\circ}$C. All experiments are conducted at 50$^{\circ}$C unless stated otherwise.
}

\subsection{COTS DRAM Chip Characterization}
\label{subsec:char_nand}
We characterize the reliability of \iey{3}{AND, NAND, OR, and NOR} operations in SK Hynix chips. While we test all three major manufacturers, we note that we do \emph{not} observe \iey{3}{AND, NAND, OR, and NOR} operations in Samsung and Micron chips.\footref{fn:mic-sam}

\begin{figure}[ht]
\centering
\includegraphics[width=\linewidth]{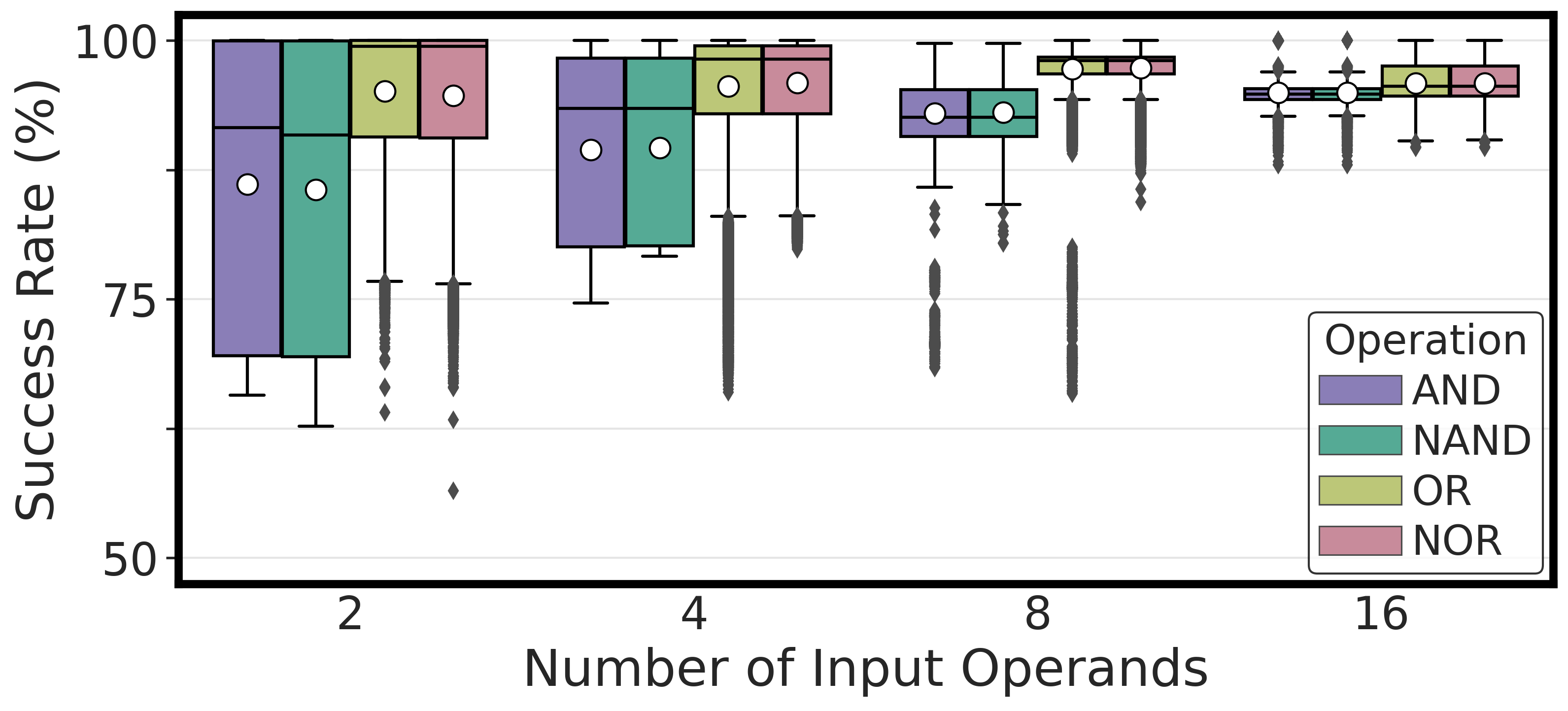}
\caption{\iey{1}{Success rates of \iey{3}{AND, NAND, OR, and NOR} operations in COTS DRAM chips with varying} numbers of input operands.}
\label{fig:and_1}
\end{figure}

\noindent\textbf{\iey{2}{N}umber of \om{3}{I}nput \om{3}{O}perands.} We evaluate the reliability of \iey{1}{\iey{3}{AND, NAND, OR, and NOR} operations with \param{2, 4, 8, and 16} input operands. \figref{fig:and_1} shows the \iey{0}{success rate} distribution across DRAM cells in all tested SK Hynix DRAM chips in a box and whiskers plot as we vary the number of input operands from 2 to 16.} \iey{2}{We make Observation\om{3}{s} 10-13 from \figref{fig:and_1}.}

\observation{\iey{2}{We can perform \iey{3}{2-input, 4-input, 8-input, and 16-input AND, NAND, OR, and NOR} operations with a very high success rate\om{3}{s} in COTS DRAM chips.}}

COTS DRAM chips can perform \iey{3}{AND, NAND, OR, and NOR} operations with 2, 4, 8, and 16 input operands with a high success rate. \iey{2}{For example, we can perform 16-input \iey{3}{AND, NAND, OR, and NOR operations} in COTS DRAM chips with average \iey{1}{success rates} of~$94.94\%$, $94.94\%$, $95.85\%$, and $95.87\%$, respectively.}

\observation{The \iey{0}{success rate} \om{3}{of bitwise operations} consistently \iey{1}{increases} as the number of input operands increases.}

For instance, \iey{1}{16-input AND operations achieve \param{10.27}\% higher average success rate than 2-input AND operations.} This trend is consistent for all tested bitwise operations. 

\observation{OR \iey{1}{(NOR)} operations have higher \iey{0}{success rate} than AND \iey{1}{(NAND)} operations.}

For example, \iey{1}{2-input OR (NOR) operations achieve} \param{10.42}\% (\param{10.60}\%) higher average \iey{0}{success rate} than 2-input AND (NAND) operations, and 16-input OR (NOR) operations have \param{0.96}\% (\param{0.97}\%) higher average \iey{0}{success rate} than 16-input AND (NAND) operations. 

\observation{There is a small average \iey{0}{success rate} difference between \one{} AND - NAND and \two{} OR - NOR operations.}

For example, the average \iey{0}{success rate} difference between two-input AND operation\atb{2}{s} and two-input NAND operations is \param{0.50}\%, and the average \iey{0}{success rate} difference between two-input OR operation\atb{2}{s} and two-input NOR operations \param{0.40}\%. 

\begin{figure}[ht]
\centering
\includegraphics[width=\linewidth]{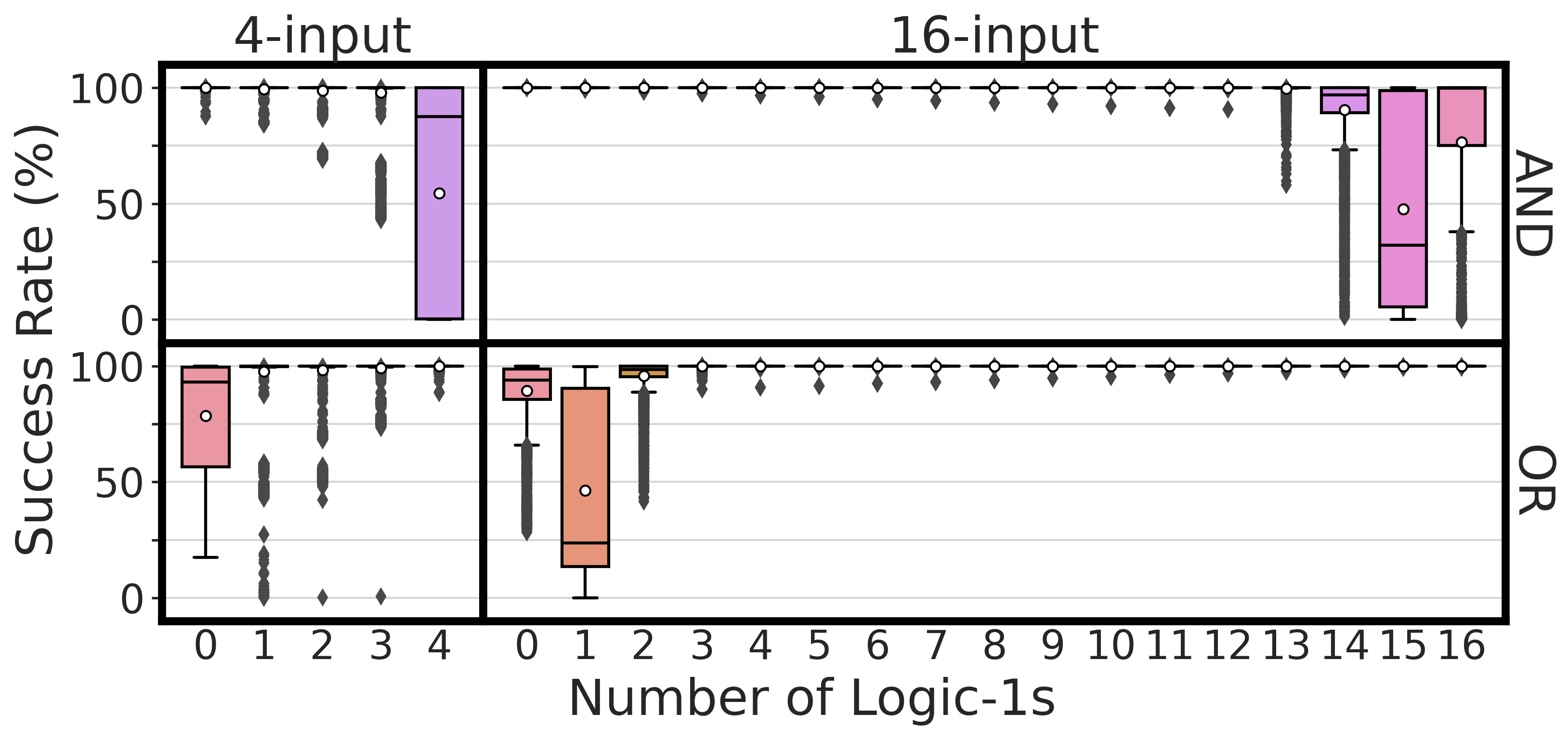}
\caption{\iey{2}{S}\iey{0}{uccess rate}\iey{2}{s} of AND and OR operations based on the number of logic-1s in the input operands.}
\label{fig:and_2}
\end{figure}

\noindent\textbf{\iey{2}{N}umber of \om{3}{L}ogic-1s in the \om{3}{Input O}perands.} \iey{1}{We analyze the impact of the number of logic-1s in the input operands on the \iey{0}{success rate} of the tested bitwise operations. To understand \atb{2}{their} impact,} we analyze 4-input AND and OR and 16-input AND and OR operations. \iey{2}{\figref{fig:and_2} shows the success rate \atb{2}{of} 4-input and 16-input AND \atb{2}{(top)} and OR \atb{2}{(bottom)} operations across all tested DRAM cells as we vary the number of logic-1s \atb{2}{(x-axis)} in their input operands. We make Observation 14 from \figref{fig:and_2}.}

\observation{The \iey{0}{success rate} of AND and OR operations heavily depends on the number of logic-1s in the input operands. AND operations experience the worst \iey{0}{success rate} when all the inputs have logic-1 or only one input has logic-0 whereas OR operations experience the worst \iey{0}{success rate} when only one \om{2}{input} or no inputs have logic-1.}

\iey{1}{For example, for the 16-input (4-input) AND operation, the average success rate decreases by \param{52.43}\% (\param{45.43}\%) when we increase the number of logic-1s in the input operands from zero to fifteen (four). Similarly, for the 16-input (4-input) OR operation, the average \iey{0}{success rate} decreases by \param{53.66}\% (\param{21.46}\%) when we decrease the number of logic-1s in the input operands from sixteen (four) to one (zero).}

\iey{1}{We hypothesize that the voltage difference in two terminals of \atb{2}{a} sense amplifier is too small for \atb{2}{the sense amplifier} to reliably amplify \atb{2}{the bitline voltage} (as sense amplifiers are \emph{not} designed \atb{2}{to reliably \om{3}{support} simultaneous multiple row activation}) \atb{2}{when the input operands of AND and OR operations are set to certain values}. For example, in \atb{2}{a} 4-input AND operation, the voltage level on reference subarray's bitline\atb{2}{s} is 3.5VDD.\footref{footnote:model} When all inputs of the AND operation are logic-1, the voltage level on each bitline in the compute subarray is 4VDD \atb{2}{and when all inputs of the AND operation are logic-0, the voltage level is} GND \atb{2}{(or 0VDD)}. As a result, the voltage difference \atb{2}{in} the two sense amplifier terminals is significantly smaller in the AND operation with all logic-1 inputs \atb{2}{(the difference is 0.5VDD)} compared to \atb{2}{another AND operation with} all logic-0 inputs \atb{2}{(the difference is 3.5VDD)}.}

\begin{figure}[htbp]
\centering
\begin{subfigure}[b]{0.49\linewidth}
    \captionsetup{skip=2pt}
    \centering
    \includegraphics[width=\linewidth]{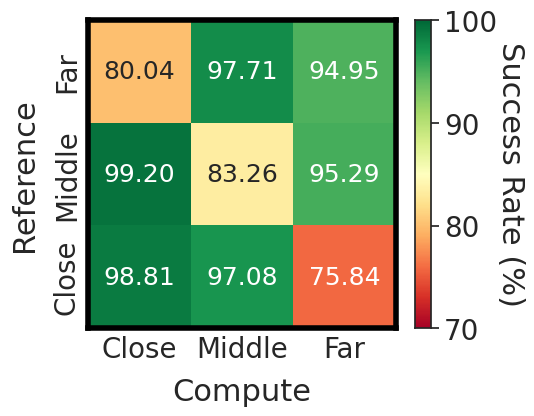}
    \caption{AND}   
    \label{subfig:ht_and}
\end{subfigure}
\hfill
\begin{subfigure}[b]{0.49\linewidth}  
    \captionsetup{skip=2pt}
    \centering 
    \includegraphics[width=\linewidth]{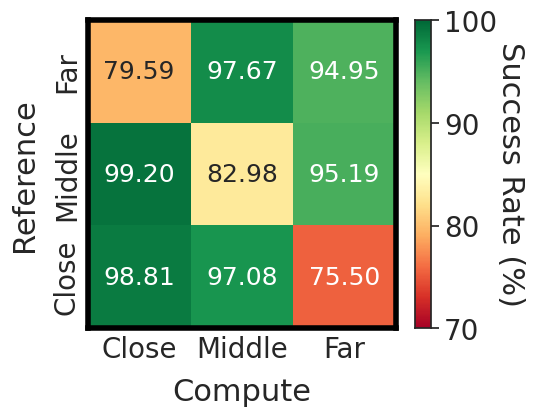}
    \caption{NAND}  
    \label{subfig:ht_nand}
\end{subfigure}

\begin{subfigure}[b]{0.49\linewidth}   
    \captionsetup{skip=2pt}
    \centering 
    \includegraphics[width=\linewidth]{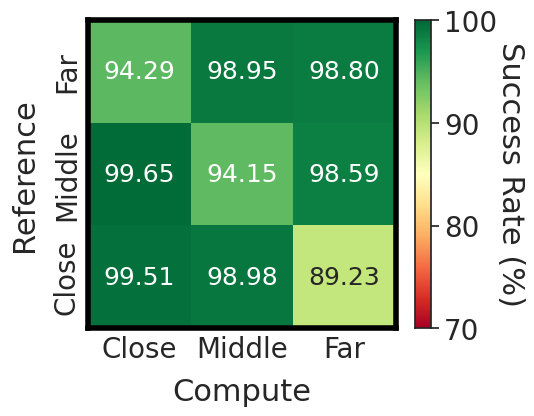}
    \caption{OR}    
    \label{subfig:ht_or}
\end{subfigure}
\hfill
\begin{subfigure}[b]{0.49\linewidth}
    \captionsetup{skip=2pt}
    \centering 
    \includegraphics[width=\linewidth]{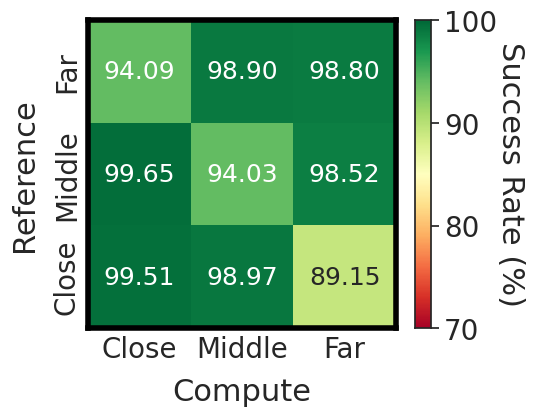}
    \caption{NOR}  
    \label{subfig:ht_nor}
\end{subfigure}
\caption{Success rate\om{2}{s} of AND, NAND, OR, and NOR operations based on the distance of activated rows to sense amplifiers.} 
\label{fig:and_4}
\end{figure}

\noindent\textbf{\om{2}{D}istance to the \om{3}{S}ense \om{3}{A}mplifier.} \figref{fig:and_4} shows the average
success rates of AND, NAND, OR, and NOR operations (i.e., the mean success rate of every tested DRAM cell \atb{2}{for} all tested numbers of input operands) using a heatmap plot based on the distance of all simultaneously activated rows in the compute subarray (y-axis) and all simultaneously activated rows in the reference subarray (x-axis) to the sense amplifiers that physically reside between (i.e., shared by) the compute subarray and the reference subarray. \iey{2}{We make Observation 15 from \figref{fig:and_4}.}

\observation{\iey{1}{The success rates of AND, NAND, OR, and NOR operations significantly vary based on the distance of the simultaneously activated rows to the sense amplifiers.}}

\iey{1}{The physical location of activated rows can lead to variations in the average \iey{0}{success rate} of up to \param{23.36}\% for AND, \param{23.70}\% for NAND, \param{10.42}\% for OR, and \param{10.50}\% for NOR operations. We hypothesize that the large variation in the success rate can indicate a strong influence of \emph{design-induced variation}~\cite{lee2017design}. Design-induced variation causes cell access latency characteristics to vary deterministically based on a cell’s physical location in the memory chip (e.g., its distance to the sense amplifier).}

\begin{figure}[ht]
\centering
\includegraphics[width=\linewidth]{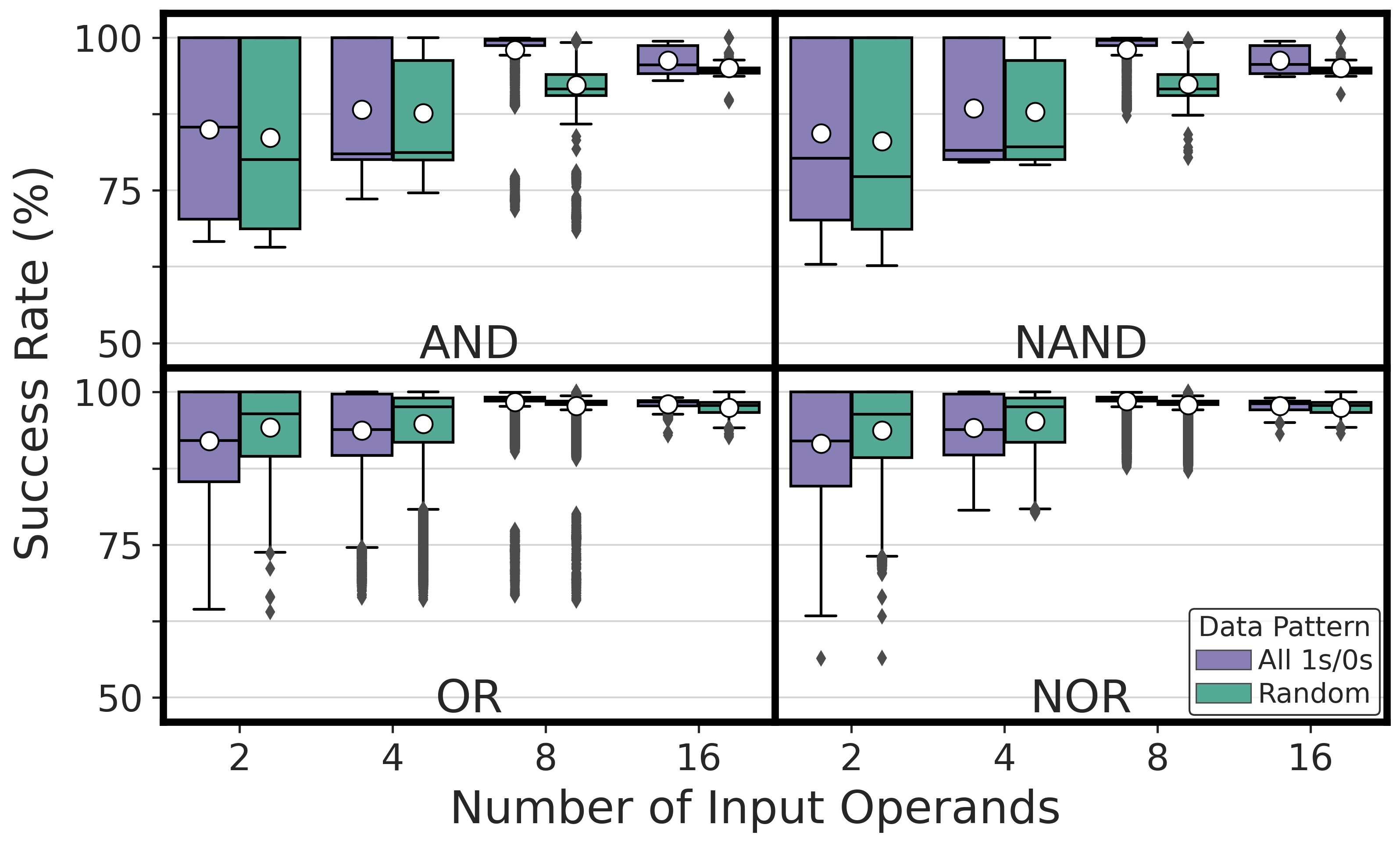}
\caption{\iey{0}{Success rates} of AND, NAND, OR, and NOR operations with different data patterns.}
\label{fig:and_3}
\end{figure}

\noindent\textbf{Data \om{3}{P}attern.} We analyze the effect of data pattern on the \iey{0}{success rate} for each operation as we vary the number of input operands \iey{2}{2} from to 16. \figref{fig:and_3} shows \iey{1}{the success rate distribution across all tested DRAM cells for two data patterns: the all 1s/0s data pattern and the random data pattern}. \iey{2}{We make Observation 16 from \figref{fig:and_3}.}

\observation{Data pattern slightly affects the \iey{0}{success rate} of AND, NAND, OR, and NOR operations.}

For example, AND operations with random data patterns have \param{1.43}\% lower average \iey{0}{success rate} than AND operations with all-1s/0s across every tested number of input operands. \atb{2}{NAND, OR, and NOR operations also have higher success rates with all-1s/0s than with random data patterns.} 
\iey{2}{We hypothesize that the variation in success rate across tested data patterns could be caused by \emph{parasitic coupling capacitance between adjacent
bitlines}~\cite{al2004effects,liu2013experimental,lee2010mechanism,nakagome1988impact,redeker2002investigation,khan2016parbor}. Prior works~\cite{al2004effects,liu2013experimental,lee2010mechanism,nakagome1988impact,redeker2002investigation,khan2016parbor} show the effect of parasitic coupling capacitance between adjacent bitlines: adjacent cells can disturb each other depending on the values stored in them, which can result in a failure during charge sharing or sensing and amplification operations.}

\begin{figure}[ht]
\centering
\includegraphics[width=\linewidth]{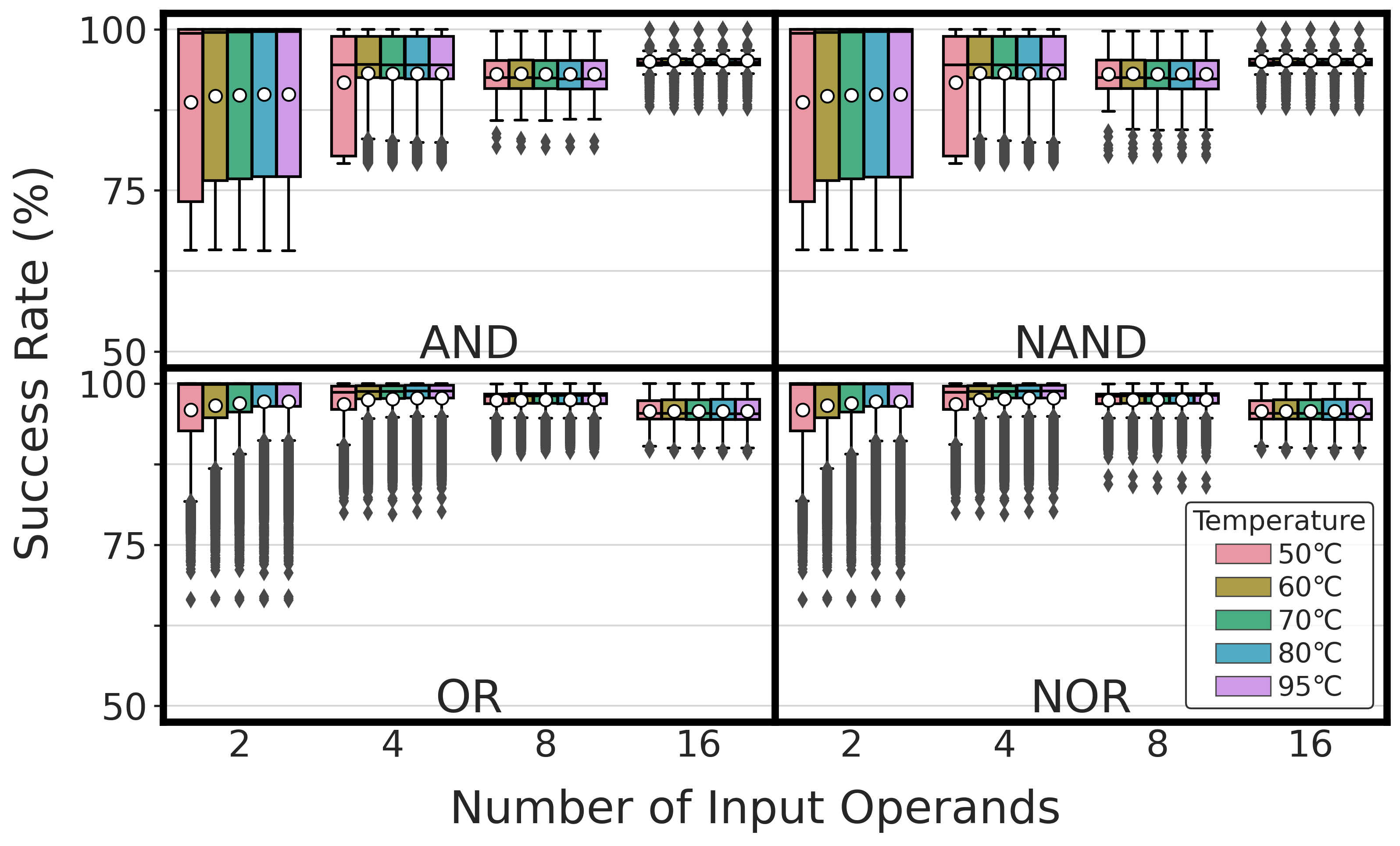}
\caption{Success rates of AND, NAND, OR, and NOR operations at different DRAM chip temperatures.}
\label{fig:and_temp}
\end{figure}

\noindent\textbf{Temperature.} \figref{fig:and_temp} shows the \iey{1}{success rate distribution of AND, NAND, OR, and NOR operations at five different temperature levels: 50$^{\circ}$C, 60$^{\circ}$C, 70$^{\circ}$C, 80$^{\circ}$C, and 95$^{\circ}$C. The figure clusters boxes into six groups based on the number of input operands. In
each cluster of boxes, temperature increases from left to right.} \iey{2}{We make Observation 17 from \figref{fig:and_temp}.}

\observation{Temperature has a small effect on the \iey{0}{success rate} of AND, NAND, OR, and NOR operations in COTS DRAM chips.}

\iey{2}{When the DRAM chip temperature increases from 50$^{\circ}$C to 95$^{\circ}$C, the highest variation (across all x-axis values) in the average success rate of} AND, NAND, OR, and NOR operations are \iey{2}{1.66\%, 1.65\%, 1.63\%, and 1.64\%}, respectively.

\takeaway{COTS DRAM chips can perform up to \param{16}-input AND, NAND, OR, and NOR \atb{2}{Boolean} operations \om{3}{(at very high success rates)}. These \atb{2}{Boolean} operations are highly resilient to temperature changes.}

\begin{figure}[ht]
\centering
\includegraphics[width=\linewidth]{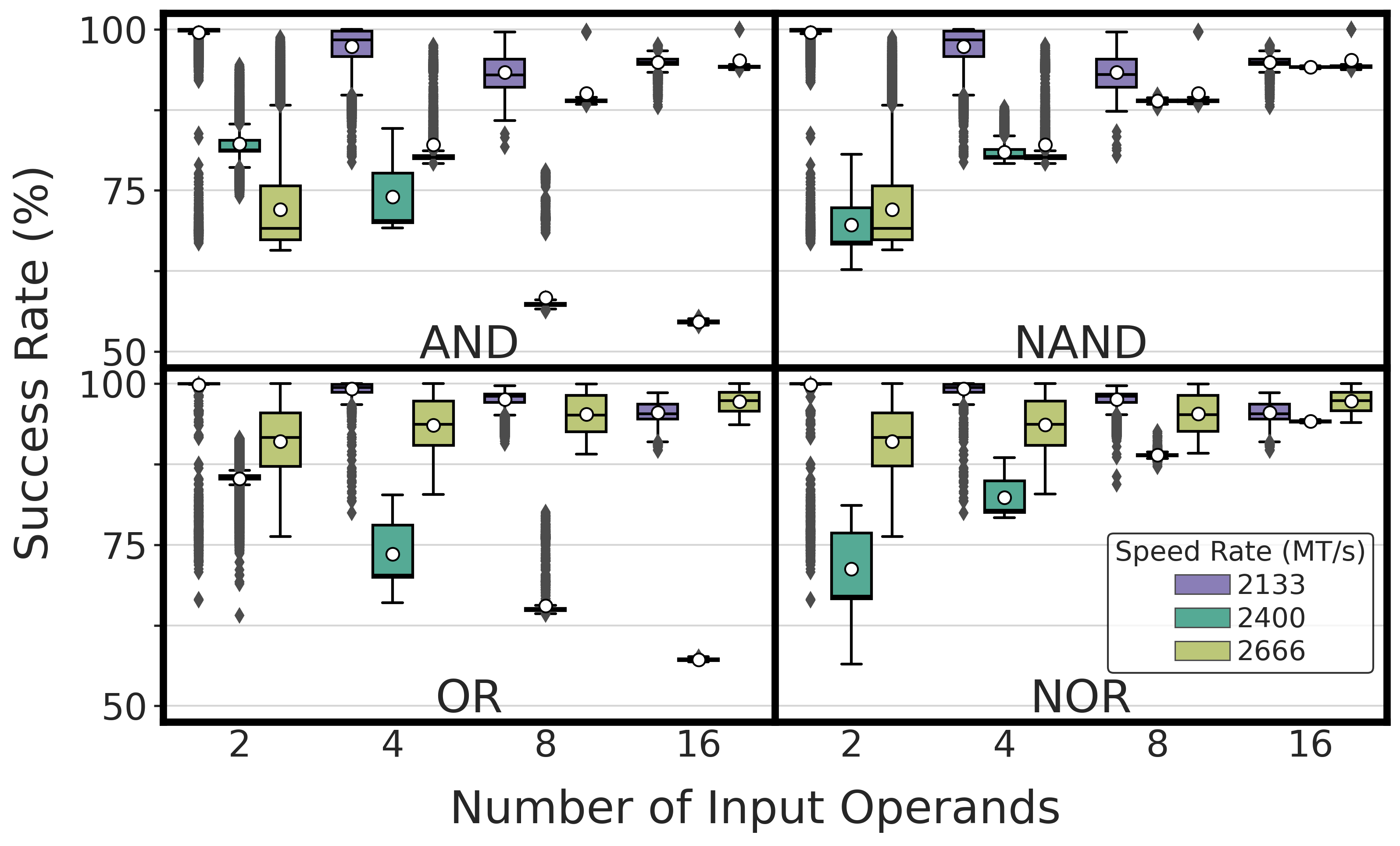}
\caption{Success rates of AND, NAND, OR, and NOR operations for three DRAM speed rates.}
\label{fig:speed_and}
\end{figure}

\noindent\textbf{DRAM \om{3}{S}peed \om{3}{R}ate.}
We analyze the effect of DRAM speed rate on the \iey{0}{success rate} of AND, NAND, OR, and NOR operations. \figref{fig:speed_and} shows the \iey{0}{success rate} distribution of AND, NAND, OR, and NOR operations \iey{1}{across DRAM cells \atb{2}{where} the x-axis \atb{2}{shows} the number of input operands rows and the hue \atb{2}{of a bar shows} the DRAM speed rate. In each cluster of boxes, the DRAM speed rate increases from left to right.} \iey{2}{We make Observation 18 from \figref{fig:speed_and}.}

\observation{The DRAM speed rate significantly affects the \iey{0}{success rate} of AND, NAND, OR, and NOR operations.}

For example, for \iey{3}{the 4-input} NAND operation, the \iey{0}{success rate} reduces by 29.89\% when the DRAM speed rate increases from 2133 MT/s to 2400 MT/s. 

\begin{figure}[ht]
\centering
\includegraphics[width=\linewidth]{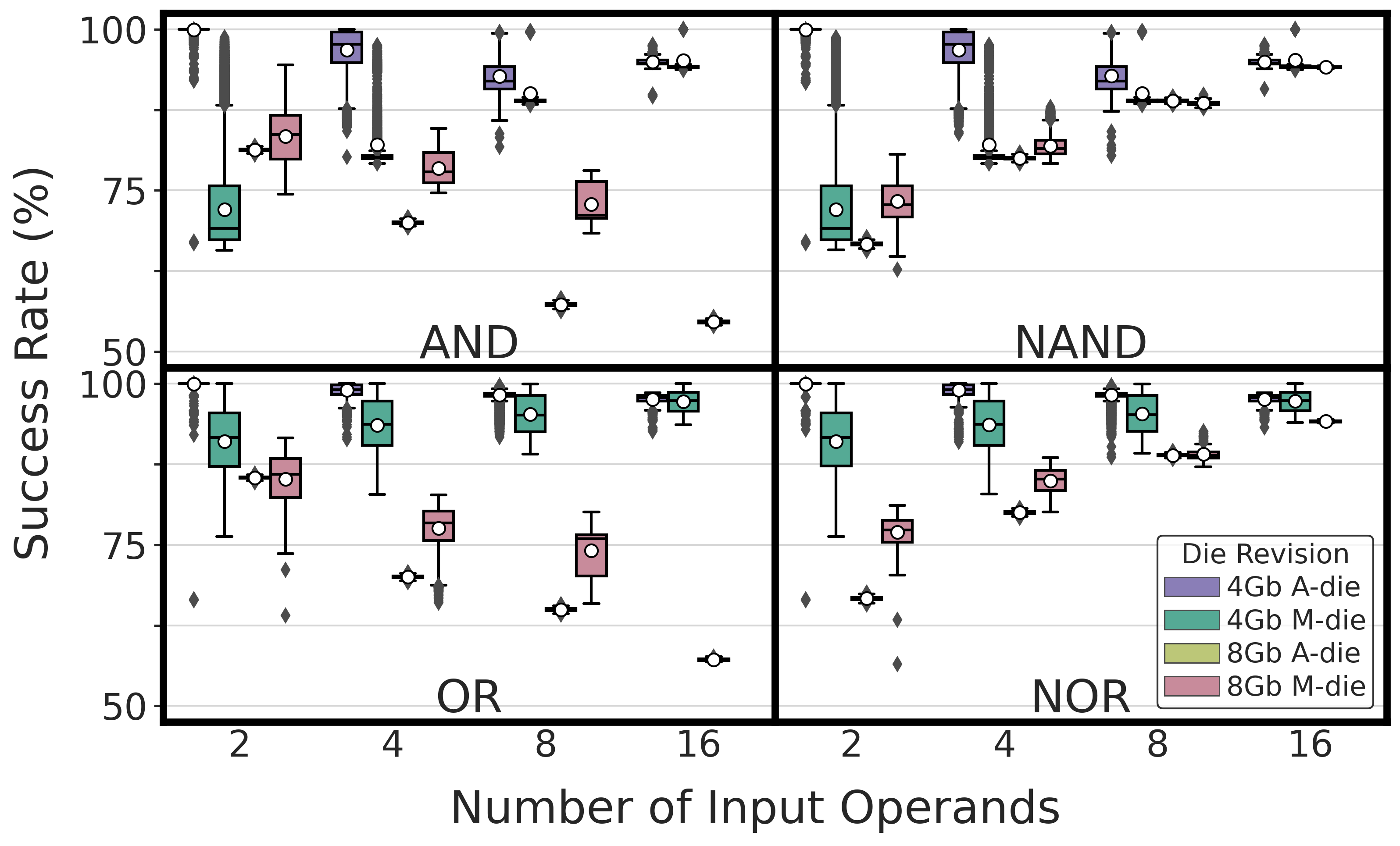}
\caption{\iey{1}{Success rates of many-input AND, NAND, OR, and NOR operations for different \iey{2}{chip density and die revision combinations}.}}
\label{fig:die_and}
\end{figure}

\noindent\textbf{\om{2}{Chip \om{3}{D}ensity and \om{3}{D}ie \om{3}{R}evision.}} 
\iey{1}{We analyze \iey{2}{the effect of chip density and die revision on} the reliability of many-input AND, NAND, OR, and NOR operations. To do so, we use SK Hynix DRAM modules spanning different \om{2}{chip densities and die revisions}: 4Gb A-die, 4-Gb M-die, 8Gb A-die, and 8Gb M-die.\footref{fn:die} \figref{fig:die_and} shows the success rate distribution across DRAM cells, with the x-axis representing the module’s \om{2}{chip density and die revision}.\footnote{We note that the tested 8Gb M-die SK Hynix module can perform up to 8-input bitwise operations as we observe that we can simultaneously activate up to 16 rows in neighboring subarrays (i.e., 8:8 activation). We provide every tested DRAM module's computational capability in the extended version of this paper~\cite{yuksel2024functionally}.}} \iey{2}{We make Observation 19 from \figref{fig:die_and}.}

\observation{\iey{1}{\iey{2}{Chip density and die revision affect the} success rate of many-input AND, NAND, OR, and NOR operations.}}

\iey{1}{For example, the average success rate of the two-input AND operation decreases by 27.47\% from 4Gb A-die to 4Gb M-die, whereas the two-input AND operation's average success rate increases by 2.11\% from 8Gb A-die to 8Gb M-die.}

\takeaway{\iey{1}{The reliability of many-input NAND, NOR, AND, and OR operations significantly varies across DRAM \iey{2}{chips with different} speed rates, die revisions, and chip densities.}}

\setcounter{version}{3}
\section{\iey{2}{Limitations of Tested COTS DRAM Chips}}
\label{sec:discussion}
We \om{3}{demonstrated} that COTS DRAM chips are fundamentally capable of performing functionally-complete Boolean operations \om{3}{using \nModule{}} DRAM modules from two major manufacturers\om{3}{,} \om{2}{and provided underlying hypotheses of why we observe \om{3}{such computation capability} based on \om{3}{the} operational principles of modern DRAM} \om{3}{chips}. 

In this section, we discuss \iey{2}{two key limitations of COTS DRAM chips \atb{2}{in performing functionally-complete Boolean operations}.}

\noindent\textbf{Limitation 1\om{3}{.} \atb{2}{Some COTS DRAM chips do \emph{not} support \emph{all} Boolean operations}.} 
While we test COTS DRAM chips from all three major manufacturers (i.e., SK Hynix, Samsung, and Micron), \om{2}{we report major positive results and detailed evaluations in chips from two major manufacturers,} SK Hynix and Samsung. In SK Hynix chips, we can \emph{simultaneously} and \emph{sequentially} activate multiple rows in neighboring subarrays, and thus, we can perform \om{3}{\emph{all}} the tested bitwise operations. In Samsung chips, we can \om{3}{\emph{only}} perform sequential row activation, and thus, we observe \om{2}{only the} NOT operation. In Micron chips, we do \om{3}{\emph{not}} observe simultaneous or sequential multiple-row activation in neighboring subarrays. Hence, we do not observe any bitwise operations in Micron chips. Prior work~\cite{yaglikci2022hira} proposes a hypothesis that could potentially explain Samsung and Micron chips' behavior. These \atb{2}{existing} DRAM chips ignore a DRAM command when the command greatly violates manufacturer-recommended timing parameters, i.e., the DRAM chip acts as if it did not receive the DRAM command. \om{3}{If such limitation was not imposed, we believe these DRAM chips are also fundamentally capable of performing the operations we examine in this work.}

\noindent\textbf{Limitation 2: \atb{2}{Tested COTS DRAM chips support up to \emph{only} 16-input Boolean operations}.}
We simultaneously active up to 48 rows in two neighboring subarrays, enabling us to perform up to 16-input NAND, NOR, AND, and OR operations. It is possible that other untested DRAM chips can simultaneously activate more rows, allowing them to perform more than 16-input bulk bitwise operations. \iey{2}{We hypothesize that} the number of simultaneously activated rows and the location of the activated rows depend on how the row decoder circuitry of a DRAM chip is designed. We believe that \atb{2}{making} row decoder circuitry \iey{2}{of DRAM chips} \iey{2}{publicly available} \atb{2}{could} help research\atb{2}{ers} develop a better understanding of the computational capability of COTS DRAM chips without exhaustive reverse engineering efforts.

\iey{3}{We believe and hope \atb{2}{that our work} inspires future DRAM designs that explicitly \om{2}{and reliably} support \om{2}{such operations building on} our proof-of-concept demonstration in \om{2}{existing} COTS DRAM chips.}

\setcounter{version}{3}
\section{Related Work}
\label{sec:related-work}
To our knowledge, this is the first work that demonstrates \one{} functionally-complete Boolean \atb{2}{operations} and \two{} many-input (i.e., more than two\om{2}{-}input) NAND, NOR, AND and OR operations in COTS DRAM chips. We \om{2}{discuss related works in \atb{2}{three} synergistic directions: 1) Processing-using-DRAM (PuD) in COTS DRAM chips, 2) PuD in modified DRAM chips, and 3) simultaneous multiple-row activation in two subarrays.}

\subsection{PuD in COTS DRAM Chips}

\noindent\textbf{Bulk Bitwise Operations~\cite{gao2019computedram,olgun2023dram,gao2022frac}.}
ComputeDRAM~\cite{gao2019computedram} activates three rows simultaneously (i.e., triple-row activation~\cite{seshadri2017ambit,seshadri.bookchapter17,seshadri2015fast,seshadri2016buddy,seshadri2016processing,seshadri2019dram}) in off-the-shelf DDR3 chips \atb{2}{to perform} three-input majority and two-input AND and OR operations~\cite{seshadri2017ambit,seshadri.bookchapter17,seshadri2015fast,seshadri2016buddy,seshadri2016processing,seshadri2019dram}. FracDRAM~\cite{gao2022frac} \atb{2}{shows that a DRAM cell in COTS DDR3 chips can store fractional values (e.g., VDD/2). FracDRAM uses fractional values to perform MAJ3 operations while simultaneously activating four DRAM rows in the same subarray.} 
DRAM Bender~\cite{olgun2023dram} demonstrates two\om{2}{-}input AND and OR operations in \atb{2}{COTS} DDR4 chips. \atb{2}{None of these works demonstrate functionally-complete Boolean operations or many-input Boolean operations.}

\noindent\textbf{Bulk Data Copy and Initialization~\cite{gao2019computedram,olgun2022pidram}.}~ComputeDRAM~\cite{gao2019computedram} demonstrates \atb{2}{bulk data copy operations in DRAM row granularity (i.e., the RowClone~\cite{seshadri2013rowclone} operation) in COTS DDR3 chips}. PiDRAM~\cite{olgun2022pidram} \iey{2}{provides a flexible end-to-end FPGA-based framework that enables \atb{2}{real system studies} of PuD techniques, such as RowClone~\cite{seshadri2013rowclone,gao2019computedram}.}

\noindent\textbf{Security Primitives~\cite{kim2019drange,talukder2019exploiting,olgun2021quactrng,keller2014dynamic,sutar2016d,xiong2016run,kim2018dram,hashemian2015robust,tehranipoor2016dram,eckert2017drng,tehranipoor2016robust}.}~Prior works demonstrate DRAM-based true random number generators (TRNGs)~\cite{olgun2021quactrng,kim2019drange,talukder2019exploiting,keller2014dynamic,sutar2016d,eckert2017drng,tehranipoor2016robust} and physical unclonable functions (PUFs)~\cite{keller2014dynamic,sutar2016d,xiong2016run,kim2018dram,hashemian2015robust,tehranipoor2016dram} using COTS DRAM chips. \atb{2}{We highlight one} state-of-the-art DRAM-based TRNG~\cite{olgun2021quactrng} \atb{2}{that} generates true random numbers in COTS DDR4 chips by simultaneously activating four rows \om{2}{in the same subarray}. Our key observation, simultaneously activating multiple rows in \om{3}{\emph{neighboring}} subarrays, \atb{2}{could} also be leveraged to generate true random numbers.

\subsection{PuD in Modified DRAM Chips} 
Various prior works~\cite{li2017drisa,Li2018SCOPEAS,seshadri2016buddy,seshadri2015fast,seshadri2016processing,seshadri2017ambit,seshadri2019dram,wu2022dram,xin2019roc,seshadri.bookchapter17,seshadri2013rowclone,xin2020elp2im,seshadri2018rowclone,ferreira2022pluto,deng2018dracc,besta2021sisa,angizi2019graphide,deng2019lacc,sutradhar2021look,sutradhar2020ppim,zhou2022flexidram, wang2020figaro,deoliveira2024mimdram} enable bulk operations in DRAM chips by \om{3}{\emph{modifying}} the DRAM circuitry. We demonstrate that COTS DRAM chips are capable of performing functionally-complete Boolean operations. \om{2}{We believe truly supporting operations that we demonstrate in DRAM requires \om{3}{proper} modifications to DRAM circuitry and standards. Yet, our demonstration shows that existing COTS DRAM chips are already quite capable of computation, and such modifications to DRAM are \om{3}{very promising and} likely to be \om{3}{very} fruitful.}

\subsection{Simultaneously Activating Two Rows in \\Two Different Subarrays in COTS DRAM Chips} 
A prior work (HiRA)~\cite{yaglikci2022hira} demonstrates \atb{2}{that real DRAM chips are capable of activating} two rows (in quick succession) in electrically isolated (i.e., \emph{not} physically adjacent) subarrays (\om{3}{called \emph{hidden row activation}}). \atb{2}{This work uses hidden row activation to parallelize a DRAM row's refresh operation with refresh or activation of other rows in the same bank.}
\setcounter{version}{3}
\section{Conclusion}
\label{sec:conclusion}
In this work, we experimentally demonstrate that COTS DRAM chips can perform functionally-complete Boolean operations (i.e., NOT and AND/OR, NAND, and NOR) and many-input (i.e., up to 16-input) NAND, NOR, AND and OR operations. Through an extensive characterization using \nChip{} COTS DDR4 DRAM chips from \nModule{} DRAM modules, we show that COTS DRAM chips can execute NOT and many-input NAND, NOR, AND, and OR operations with high reliability, and data pattern and temperature changes only slightly affect their reliability. We believe that \om{2}{our empirical} results demonstrate the potential of using DRAM as a powerful computation substrate. \iey{2}{We hope future works build upon our comprehensive study to better characterize\om{3}{,} understand\om{3}{, and enhance} the computational capability of DRAM chips.}

\section*{Acknowledgements}
We thank the anonymous reviewers of HPCA 2024 for their encouraging feedback. 
We thank the SAFARI Research Group members for providing a stimulating intellectual environment. We acknowledge
the generous gifts from our industrial partners\iey{1}{, including} Google, Huawei, Intel, and Microsoft. This work is supported in part by the Semiconductor Research Corporation~\iey{2}{(SRC), the ETH Future Computing Laboratory (EFCL), and the AI Chip Center for Emerging Smart Systems (ACCESS)}.


\setstretch{0.997}
\balance 
{
  \bstctlcite{IEEEexample:BSTcontrol}
  \let\OLDthebibliography\thebibliography
  \renewcommand\thebibliography[1]{
    \OLDthebibliography{#1}
    \setlength{\parskip}{0pt}
    \setlength{\itemsep}{0pt}
  }
  \bibliographystyle{IEEEtran}
  \bibliography{refs}
}

\end{document}